\documentclass[useAMS,usenatbib]{mn2e}

\usepackage{epsfig}
\usepackage{myaasmacros}
\usepackage{amsmath}
\usepackage{Arial}
\usepackage{ulem}

\title[Magnetic fields in spiral galaxies]
{Magnetic field structure due to the global velocity field in spiral galaxies}
\author[H. Kotarba, H. Lesch, K. Dolag, T. Naab, P. H. Johansson \& F. A. Stasyszyn]{H. Kotarba$^{1,3}$\thanks{E-mail:
kotarba@usm.lmu.de}, H. Lesch$^{1}$, K. Dolag$^{2}$, T. Naab$^{1}$, P. H. Johansson$^{1}$ \& F. A. Stasyszyn$^{2}$\\
$^{1}$University Observatory Munich, Scheinerstr.1, D-81679 Munich, Germany\\
$^{2}$Max Planck Institute for Astrophysics, Karl-Schwarzschild-Str. 1, D-85741 Garching, Germany\\
$^{3}$Max Planck Institute for Extraterrestrial Physics, Giessenbachstrasse, D-85748 Garching, Germany}

\begin{document}

\date{Accepted . Received ; in original form 2008 November 14}

\pagerange{\pageref{firstpage}--\pageref{lastpage}} \pubyear{2002}

\maketitle

\label{firstpage}

\begin{abstract}
We present a set of global, self-consistent \textsl{N}-body/SPH simulations of the dynamic evolution of galactic discs with gas and including magnetic fields. We have implemented a description to follow the evolution of magnetic fields with the ideal induction equation in the SPH part of the $\textsc{Vine}$ code. Results from a direct implementation of the field equations are compared to a representation by Euler potentials, which pose a $\nabla\cdot \textbf{B}$-free description, an constraint not fulfilled for the direct implementation. All simulations are compared to an implementation of magnetic fields in the $\textsc{Gadget}$ code which includes also cleaning methods for $\nabla\cdot \textbf{B}$.

Starting with a homogeneous seed field we find that by differential rotation and spiral structure formation of the disc the field is amplified by one order of magnitude within five rotation periods of the disc. The amplification is stronger for higher numerical resolution. Moreover, we find a tight connection of the magnetic field structure to the density pattern of the galaxy in our simulations, with the magnetic field lines being aligned with the developing spiral pattern of the gas.
Our simulations clearly show the importance of non-axisymmetry for the evolution of the magnetic field.

\end{abstract}


\section{Introduction}

\begin{figure}
\begin{center}
  \epsfig{file=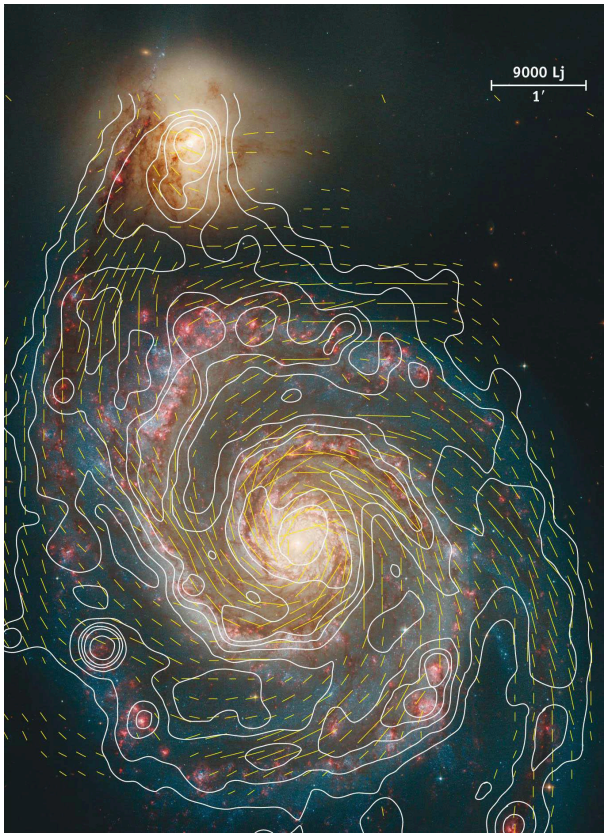, width=0.45\textwidth}
  \caption{Optical image of M51 (Hubble) overlayed with contours of total synchrotron intensity as measure for the total magnetic field (combined observations at Effelsberg and VLA at 6 cm) and vectors of magnetic field. From A. Fletcher \& R. Beck (MPIfR) and Hubble Heritage Team (STScI), published by 'Sterne und Weltraum`, September 2006.}
  {\label{M51}}
\end{center}
\end{figure}

Radio observations have revealed that disc galaxies are permeated by large scale magnetic fields ordered on kpc scales and beyond (\citealp{Beck&Hoernes1996}, \citealp{Hummel&Beck1995}, \citealp{Beck&Krause1985}). The typical field strength, determined from polarization, Faraday rotation and energy equipartition is of the order of 10 $\mu$G (e.g. \citealp{Beck2004}). The spatial structure of the $\textbf{B}$-field reflects the spiral and/or barred structure of the gas distribution within the galactic discs (\citealp{Beck2008}).
For example, Fig. \ref{M51} shows optical observations of the spiral galaxy M51 overlayed with contours of total synchrotron intensity (tracing the total magnetic field) and magnetic field vectors. It reveals the tight connection of magnetic field with the gas distribution in the galactic disc.

The motion of the gas within the gravitational potential of a galaxy strongly influences the strength and direction of the magnetic field in the interstellar medium. This can be seen by inspecting the well known induction equation of magnetohydrodynamics (MHD), i.e. the temporal evolution equation for the magnetic field,
\begin{eqnarray}
\frac{\partial\textbf{B}}{\partial t}=\nabla\times(\textbf{v}\times\textbf{B})-(\nabla\times\eta(\nabla\times\textbf{B})),\label{induktion1}
\end{eqnarray}
where $\textbf{v}$ denotes the gas velocity and $\eta$ represents the magnetic diffusivity which is inversely proportional to the electrical conductivity.

Apparently, within the frame of MHD, the role of the galaxy as a whole is simply to provide for the gas velocity field. Since the conductivity of the interstellar medium is very high, the magnetic field is closely coupled to the gas motion. It is this `frozen-in'-property of both, the magnetic field and the gas, which determines the spatial structure of the magnetic field. In other words, a detailed investigation of the velocity field of the interstellar gas in disc galaxies is necessary for a deeper physical understanding of the evolution of galactic magnetic fields.

The gas in the disc rotates differentially within the global gravitational potential. Angular momentum transport via spiral arms, bars and gravitational interaction forces the gas to move towards the central regions, and eventually, star formation activity in the disc (superbubbles, winds etc.) drives gas perpendicular to the plane of the disc towards the galactic halo. In general, the axisymmetric rotation velocity is the dominant component, followed by non-axisymmetric and radial components. The velocity components perpendicular to the disc are typically the smallest. Altogether, $\textbf{v}$ in Eq. \ref{induktion1} represents a complex three-dimensional non-axisymmetric velocity field strongly coupled to the global properties of the galaxy, including the dark matter halo, stellar component and internal disc activity.

Beside the large scale components of the gas velocity field there are also small scale velocity fluctuations of interstellar gas driven by all kinds of local disc activity, i.e. stellar winds, supernova explosions, cloud-cloud collisions, galactic winds, etc (see e.g. \citealp{Ferriere1992}, \citealp{Efstathiou2000}, \citealp{Johansson&Efstathiou2006}, \citealp{Kulsrud&Zweibel2008}, \citealp{Gressel2008}). These unordered velocity components generate two effects which are known as helicity  (in terms of a convective turbulent motion perpendicular to the disc) and turbulent diffusion (magnetic field lines with antiparallel direction reconnect and annihilate partially). Helicity supports the amplification of the magnetic field, whereas turbulent diffusion reduces the field strength (see, e.g. \citealp{Brandenburg2005} for a review of nonlinear dynamo theory). Therefore, an incorporation of these small scale velocity components into the analysis requires some manipulation of the induction equation (Eq. \ref{induktion1}) in terms of a mean-field theory (\citealp{Steenbeck&Krause1969}, \citealp{Wielebinski&Krause1993}, \citealp{Sur&Shukurov2007}). Within the frame of the mean-field description the velocity and magnetic fields are considered as superpositions of the mean and fluctuating parts ($\textbf{v}=\langle\textbf{v}\rangle+\textbf{v}'$ and $\textbf{B}=\langle\textbf{B}\rangle+\textbf{B}'$). The fluctuating velocity components are coupled to small-scale fluctuations of the magnetic field. The coupling terms are then given by $\nabla\times\alpha\langle\textbf{B}\rangle$, where $\alpha=\frac{1}{3}\tau\langle\textbf{v}'\cdot(\nabla\times\textbf{v}')\rangle$ (\citealp{Zeldovich}), and by $\eta_T\Delta \langle\textbf{B}\rangle$, where $\eta_T$ now describes the turbulent diffusion coefficient $\eta_T\propto v_\mathrm{turb}\cdot l_\mathrm{turb}$, where $v_\mathrm{turb}$ and $l_\mathrm{turb}$ are the typical velocity and length scale of the turbulent motion, respectively.

This leads to the dynamo equation
\begin{eqnarray}
\frac{\partial\textbf{B}}{\partial t}=\nabla\times(\textbf{v}\times\textbf{B})+\nabla\times\alpha\textbf{B},\label{alphadynamo}
\end{eqnarray}
where we have neglected the diffusivity and dropped the mean-brackets for convenience (here, and in the following, $\textbf{B}$ and $\textbf{v}$ refer to their mean values).

Eq. \ref{alphadynamo} is the central equation of cosmic mean field dynamos. It describes the circle of amplification of the different components. The classical dynamo model describes the amplification of the magnetic field through the following chain of $\alpha$ (convective turbulence) and $\Omega$ (differential rotation) actions: The radial component $B_r$ is amplified through $\alpha$-action from turbulence; then $B_\varphi$ is generated from $B_r$ through $\Omega$-action from the shear of the galactic differential rotation. Such an $\alpha\Omega$ mean field dynamo amplifies the magnetic field by repeating the chain of $\alpha$ and $\Omega$ actions (see \citealp{Widrow2002} and \citealp{Stefani2008} for a review of dynamo theory). However, the origin of the $\alpha$-effect is still under discussion (\citealp{Cattaneo&Vainshtein1991}, \citealp{Vainshtein&Cattaneo1992}, \citealp{Kulsrud&Anderson1992}).

We emphasize that the described classical dynamo models use only one velocity component, the differential rotation. To be more precise the role of any deviation from axisymmetry is considered to be unimportant for the evolution of the large-scale magnetic field, which is not necessarily true in real galaxies.

On this account, there have been three-dimensional numerical simulations using an analytical turbulent velocity field, where deviations from axisymmetry were incorporated in the gas- and turbulence-profiles (\citealp{Rohde&Elstner&Rudiger1997}, \citealp{Rohde&Elstner1998}).
These studies showed, that even accounting for the $\alpha$-effect calculated out of the analytical velocity field an initial magnetic field cannot survive for more than 500 Myr.

Moreover, \citealp{Elstner&Otmianowska-Mazur2000} performed \textsl{N}-body simulations of two component (collisionless stars and gaseous clouds moving in the gravitational potential of the stellar population), self-gravitating discs embedded in an analytical bulge- and halo-potential. These simulated clouds provided an already very good approximation of the gas velocity field. However, full hydrodynamics was not incorporated. The obtained velocity field was used in an $\alpha\Omega$-dynamo description. Without including the $\alpha$-effect, the non-azimuthal 3D gas flow alone did not provide an amplification of the magnetic field. The field got amplified by several orders of magnitude within 0.7 Myr only when the $\alpha$-effect was included. In addition, they found an alignment of the magnetic field with the developed spiral pattern of the disc.

Recently, \citet{Dobbs&Price2008} performed three dimensional, full MHD, single and two-component (cold and hot gas) simulations using smooth particle hydrodynamic (SPH) methods to treat MHD. They applied a spiral potential to the gas, thus, the self-induced formation of spiral structure was not included. Their work concentrated on structure formation in the disc, like molecular clouds and inter-arm spurs. They found that the
main effect of adding a magnetic field to these calculations was to inhibit the formation
of structure in the disc. They did not consider global enhancement and structure formation of the magnetic field, but nevertheless, they found that the global magnetic field was following the large scale velocity field.

It is the aim of this paper to present further steps towards a more complete dynamo model. We perform for the first time a set of self-consistent \textsl{N}-body calculations of a spiral galaxy including hydrodynamics as well as the induction equation via the SPH method to obtain the complex three dimensional velocity field. Compared to all previous work, we use no analytical potential for any component of the galaxy. All components (disc, gas, bulge and halo) are represented by particles which are treated as self-gravitating \textsl{N}-body-particles, while hydrodynamics is applied to the gas component only. We use more than one order of magnitude more particles than \citet{Elstner&Otmianowska-Mazur2000}. We follow the evolution of the magnetic field according to the induction equation (eq. \ref{induktion1}). Thus, we have implemented the SPH variant of the induction equation as well as the representation of the magnetic fields by Euler potentials in the SPH code $\textsc{Vine}$ and compare the results with simulations performed using the SPH code $\textsc{Gadget}$. \textsl{N}-body/SPH methods are well adapted for simulating whole galactic discs as the simulated discs stay stable for at least 15 dynamical times (where we define the dynamical time for a disc galaxy as its half mass rotation period). As we will show in section \ref{SIMULATIONS}, our discs are forming spiral structure without applying a spiral potential or any other mechanism to provide extraordinary flows.

In summary, we investigate here the kinematic reaction of a large-scale magnetic field on the complete three-dimensional, large-scale velocity field of a disc galaxy obtained from the \textsl{N}-body SPH simulations, using two different numerical codes.

The paper is organized as follows: Section \ref{THEORY} gives shortly the theory of magnetic field evolution in differentially rotating systems.
A summary of the SPH method and the treatment of magnetic fields including the method based on Euler potentials is given in section \ref{NUMERICS}. The simulations together with a comparison of the performance of the $\textsc{Vine}$ and $\textsc{Gadget}$ codes are presented in section \ref{SIMULATIONS}.
The results are discussed in section \ref{DISCUSSION}, where we also analyse the terms of the induction equation in detail.
Finally, we summarise and conclude in section \ref{SUMMARY}.

\section{Theoretical Expectations}\label{THEORY}

When only studying the effect of the gas velocity on the evolution of the magnetic field, we can neglect the diffusive term in eq. \ref{induktion1}. Keeping this term, one would physically except the magnetic field to dissipate depending on the value of $\eta$ and reconnect when converse magnetic field lines come together. Technically, $\eta$ is not always assumed to be spatially dependent, so that the diffusive term reads $-\eta\nabla^2\textbf{B}$. However, this formulation leads only to an effective smoothing, and not a real diffusion of the magnetic field. Neglecting the diffusive term thus corresponds to considering an upper limit of field amplification. Additionally, $\eta$ is assumed to be small except within strong shocks.

The induction equation \ref{induktion1} then yields
\begin{eqnarray}
\frac{\partial\textbf{B}}{\partial t}&=&(\textbf{B}\cdot\nabla)\textbf{v}+\textbf{v}\underbrace{(\nabla\cdot\textbf{B})}_{=0}
-\textbf{B}(\nabla\cdot\textbf{v})-(\textbf{v}\cdot\nabla)\textbf{B},\label{induktionsglg}
\end{eqnarray}
and applying cylindrical coordinates eq. \ref{induktionsglg} reads:
\begin{eqnarray}
\frac{\partial B_r}{\partial t}&=&
-B_r\frac{v_r}{r}-\frac{1}{r}B_r\frac{\partial v_\varphi}{\partial\varphi}-B_r\frac{\partial v_z}{\partial z} +\frac{1}{r}B_\varphi\frac{\partial v_r}{\partial\varphi}\notag\\
&& + B_z\frac{\partial v_r}{\partial z}
-v_r\frac{\partial B_r}{\partial r}-\frac{v_\varphi}{r}\frac{\partial B_r}{\partial\varphi}-v_z\frac{\partial B_r}{\partial z},\\
\frac{\partial B_\varphi}{\partial t}&=&
-B_\varphi\frac{\partial v_r}{\partial  r}-B_\varphi\frac{\partial v_z}{\partial z} + B_r\frac{\partial v_\varphi}{\partial r}+B_z\frac{\partial v_\varphi}{\partial z}\notag\\
&&-v_r\frac{\partial B_\varphi}{\partial r}-\frac{v_\varphi}{r}\frac{\partial B_\varphi}{\partial\varphi}-v_z\frac{\partial B_\varphi}{\partial z}-\frac{v_\varphi B_r}{r},\\
\frac{\partial B_z}{\partial t}&=&
-B_z\frac{v_r}{r}-B_z\frac{\partial v_r}{\partial r}-\frac{1}{r}B_z\frac{\partial v_\varphi}{\partial \varphi} +B_r\frac{\partial v_z}{\partial r}\notag\\
&&+\frac{1}{r}B_\varphi\frac{\partial v_z}{\partial\varphi}-v_r\frac{\partial B_z}{\partial r}-\frac{v_\varphi}{r}\frac{\partial B_z}{\partial\varphi}-v_z\frac{\partial B_z}{\partial z}.
\end{eqnarray}
These equations can be simplified to get a first idea of how magnetic fields will evolve in a galactic disc.
For a differentially rotating disc ($\partial v_\varphi /\partial r\approx 0$) with a perfectly axisymmetric velocity field, $\textbf{v}$ does not depend on $\varphi$. The same holds for an axisymmetric magnetic field. If we also assume that changes of all quantities in the $z$ direction are small compared with those in the radial direction and $B_z\simeq0$, the equations for the regular field, i.e. the field in the plane of the disc, read:
\begin{eqnarray}
\frac{\partial B_r}{\partial t}&=&
-B_r\frac{v_r}{r}\label{induktionsglg_d},\\
\frac{\partial B_\varphi}{\partial t}&=&-B_\varphi\frac{\partial v_r}{\partial  r}-\frac{v_\varphi B_r}{r}=-B_\varphi\frac{\partial v_r}{\partial  r}+B_r r\frac{\partial\Omega}{\partial r}\label{induktionsglg_e},
\end{eqnarray}
where $\Omega$ is the angular velocity.

In the absence of radial flows, the last term of eq. \ref{induktionsglg_e} describes the generation of a toroidal magnetic field from the radial component of the already present magnetic field by differential rotation. This effect is the so called $\Omega$-effect already mentioned above. Since $v_\varphi\gg (v_r,v_z)$ this term is dominant and one would expect any initial magnetic field to be first wound up by differential rotation. However, this effect alone cannot be responsible for a significant amplification of the magnetic field, as the amplification stops when all of the radial field is wound up.
However, if a gas flow in the negative radial direction (i.e. towards the centre of the disc) is present the radial field can be amplified and then be converted into a toroidal field. These radial gas flows occur when angular momentum is transported in the gas out of the disc, e.g. by spiral arms or bars.
The toroidal magnetic field can be amplified further if the radial gas flow velocity decreases with increasing galactocentric radius.

Therefore, a good understanding of the evolution of magnetic fields in galactic discs requires full information of the three-dimensional velocity field of the gas which is naturally provided by self-consistent numerical simulations. We will discuss the velocity field in our simulations and the resulting values of the different terms of the induction equation in section \ref{DISCUSSION}.

\section{Numerical methods}\label{NUMERICS}

\subsection{$\textsc{Vine}$}
The equations presented below are implemented within the OpenMP parallel
\textsl{N}-body/SPH evolution code $\textsc{Vine}$. For all details we refer the reader to \citet{VINEI} and \citet{VINEII}.

\subsubsection{SPH basics}

Within the SPH formulation a hydrodynamic quantity $A$ is interpolated by a kernel
function $W(\textbf{r}-\textbf{r}',h)$ with $\int Wd\textbf{r}=1$ and $\lim_{h\rightarrow0}W=\delta(\textbf{r}-\textbf{r}')$, where the so called smoothing length $h$ defines the spatial extent of the function $W$.
This interpolation is then discretised, so that
\begin{eqnarray}
  A_i=\sum_j m_j\frac{A_j}{\rho_j}W(\textbf{r}_i-\textbf{r}_j,h),
\end{eqnarray}
where $i$ ($j$) is the index of the particle at position $\textbf{r}_i$ ($\textbf{r}_j$) and $A_i$ ($A_j$) the value of the quantity $A$ at the position of particle $i$ ($j$). $\rho_j$ and $m_j$ denote the density and mass at position of particle $j$, respectively.

The $\textsc{Vine}$ code uses the common W4 kernel defined by \citet{Monaghan&Lattanzio1985} as
\begin{eqnarray}
W_{ij}=W(\textbf{r}_{ij},h)=\frac{\sigma}{\bar{h}^\nu_{ij}}\left\{\begin{array}{ll}
1-\frac{3}{2}\varrho^2+\frac{3}{4}\varrho^3 &  0\leq \varrho < 1 \\
\frac{1}{4}(2-\varrho)^3  & 1\leq \varrho < 2 \\
0 & \mbox{else}
\end{array}\right.,
\end{eqnarray}
where values with index $ij$ denote differences (e.g. $\textbf{r}_{ij}=\textbf{r}_i-\textbf{r}_j$) and arithmetic means (e.g. $\bar{h}_{ij}=0.5\cdot(h_i+h_j)$), respectively,  $\varrho=|\textbf{r}-\textbf{r}'|/h$, $\nu$ is the number of spatial dimensions of the system and $\sigma$ is a constant of order unity.
See \citet{Monaghan1992}, \citet{Monaghan2001basics} or \citet{Price2005SPH} for more details.

\subsubsection{Continuity equation}

As long as the kernel itself is differentiable, every function $A$ can be interpolated to a differentiable function by the procedure described above. The most common formulation of derivatives in SPH is (see e.g. \citealp{Monaghan1992}, \citealp{Price2005SPH})
\begin{eqnarray}
  (\nabla A)_i = \frac{1}{\rho_i}\sum_j m_j (A_j-A_i)\nabla_i W_{ij}.\label{gradient}
\end{eqnarray}
Using the continuity equation, the total time derivative of the density thus reads
\begin{eqnarray}
  \frac{d\rho_i}{dt}=-\rho_i(\nabla\cdot\textbf{v})_i=\sum_j m_j (\textbf{v}_i-\textbf{v}_j)\cdot\nabla_i W_{ij}.
  \label{continuum}
\end{eqnarray}

\subsubsection{Momentum and Energy equation}

A natural ansatz to derive a conservative form of the momentum equation comprising the force due to pressure gradients (in addition to the force due to the gravitational potential) is to use the Lagrange formalism together with the first law of thermodynamics. This leads to the following SPH variant of its ideal form ($\frac{d\textbf{v}}{dt}=-\frac{\nabla P}{\rho}$):
  \begin{eqnarray}
    \frac{d\textbf{v}_i}{dt}=-\frac{\nabla P_i}{\rho_i}=-\sum_j m_j
    \left(\frac{P_i}{\rho_i^2}+\frac{P_j}{\rho_j^2}\right)\nabla_i W_{ij}\label{momentum}.
  \end{eqnarray}
In this formulation momentum is conserved exactly, since the contribution of particle $j$ to the momentum of particle
$i$ is equal and negative to the contribution of particle $i$ to the momentum of particle $j$.

The change in the thermodynamical state of the gas requires an evolution equation for a state variable corresponding to the internal energy or entropy of the gas. $\textsc{Vine}$ employs an equation for the specific internal energy ($u$) of the gas. Without external heating or cooling terms, only compressional heating and cooling are important and the SPH variant of the ideal form ($\frac{du}{dt}=-\frac{P}{\rho}\nabla\cdot\textbf{v}$) reads:
\begin{eqnarray}
  \frac{du_i}{dt}=\frac{P_i}{\rho^2_i}\sum_j m_j\textbf{v}_{ij}\cdot\nabla_i W_{ij}\label{internalEnergy}.
\end{eqnarray}
To close the set of equations, an isothermal equation of state is used throughout this paper.

\subsubsection{Artificial viscosity}

Artificial viscosity is required to model shocks and angular momentum transport properly, where the latter is important to be able to simulate spiral arm formation.
The $\textsc{Vine}$ code uses the most common form of the artificial viscosity. It is described by the tensor $\mathbf{\Pi}_{ij}$ as in \cite{Monaghan1992}.

Since the value of $\mathbf{\Pi}_{ij}$ depends on the difference in velocity between the considered particles (i.e. the velocity gradient) the viscosity increases with increasing velocity gradient. Moreover, the viscosity is only applied if particles are approaching each other.

\cite{Balsara1995} suggested a viscosity limiter to avoid spurious angular momentum and vorticity transport in gas disks. However, a lower viscosity leads to a higher velocity dispersion of the gas and therefore to higher divergence of the velocity and magnetic field. As will be shown and discussed in section \ref{DISCUSSION}, this higher divergence causes a more violent magnetic field amplification.

The viscous terms within the momentum and energy equations read:
\begin{eqnarray}
\left(\frac{d\textbf{v}_i}{dt}\right)_{diss}&=&-\sum_j m_j \mathbf{\Pi}_{ij}\nabla_i W_{ij}\label{v diss}\\
  \left(\frac{du_i}{dt}\right)_{diss}&=&\frac{1}{2}\sum_j m_j \mathbf{\Pi}_{ij}\textbf{v}_{ij}\cdot\nabla_i W_{ij}\label{u_diss}
\end{eqnarray}

This treatment of viscous forces allows for a sensible description of the behaviour of gas in a spiral galaxy. However, eqs. \ref{internalEnergy} and \ref{u_diss} are not applied when using isothermal equation of state.

\subsubsection{Induction equation}

In order to follow the evolution of the magnetic field we have additionally implemented equation \ref{induktionsglg} discretised as
\begin{eqnarray}
 \frac{dB_i^\mu}{dt}&=&\frac{1}{\rho_i}\sum_j m_j[\underbrace{B_i^\mu(\textbf{v}_{ij}\cdot\nabla_i W_{ij})} _{\hat{=}-\textbf{B}(\nabla\cdot\textbf{v})} \underbrace{-v_{ij}^\mu(\textbf{B}_i\cdot\nabla_i W_{ij})}_{\hat{=}(\textbf{B}\cdot\nabla)\textbf{v}}]\notag\\
&=&\frac{1}{\rho_i}\sum_j m_j(B_i^\mu\textbf{v}_{ij}-v_{ij}^\mu\textbf{B}_i)\nabla_i W_{ij}\label{IndEqsSPH}.
\end{eqnarray}
with $\mu,\nu$ denoting the spatial directions.

\subsubsection{Euler potentials}
\label{EULER}
A well known problem related to magnetic fields within SPH is the maintenance of $\nabla\cdot \textbf{B}=0$ throughout
the simulation. Different attempts to solve this problem have been made (see \citealp{Price&Monaghan2005SPMHDIII}), examples include source term approaches (\citealp{Powell1999}) and projection methods.

Theoretically, the problem can be avoided if the magnetic field is represented by Euler potentials (\citealp{Stern1970}, \citealp{Price&Bate2007}, \citealp{Rosswog&Price2007}), an approach we have also implemented into the $\textsc{Vine}$ code.

The magnetic field is expressed as a function of two scalar potentials $\alpha_E$ and $\beta_E$ as:
\begin{eqnarray}
\textbf{B}=\nabla\alpha_E\times\nabla\beta_E \label{euler}.
\end{eqnarray}
Taking the divergence of $\textbf{B}$ we get:
\begin{eqnarray}
  \nabla\cdot\textbf{B}&=&\nabla\cdot(\nabla\alpha_E\times\nabla\beta_E)\notag\\
  &=&\nabla\beta_E\underbrace{(\nabla\times\nabla\alpha_E)}_{=0} -\nabla\alpha_E\underbrace{(\nabla\times\nabla\beta_E)}_{=0}=0.
\end{eqnarray}
Thus, the divergence constraint is fulfilled by construction.

Moreover, for the ideal case ($\eta = 0$) the Euler potentials for each particle (i.e. the convective values of the potentials) are direct tracers of the magnetic field and stay constant with time,
\begin{eqnarray}
\frac{d\alpha_E}{dt}&=&0,\\ \frac{d\beta_E}{dt}&=&0,
\end{eqnarray}
   thus one does not need to perform an additional integration when following magnetic fields, leading to a higher accuracy of the calculation. The variation of magnetic field is only due to the motion of
the particles, which corresponds to the advection of magnetic field lines by Lagrangian particles (frozen flow) (\citealp{Stern1970}).

Moreover, the formulation by Euler potentials guarantees conservation of magnetic helicity,
\begin{eqnarray}
  H=\int_V \textbf{A}\cdot \textbf{B}d^3\textbf{x},
\end{eqnarray}
which is again reasonable for ideal treatment. Actually, the magnetic helicity is zero, since for $\textbf{A}=\alpha_E\nabla\beta_E$ and equivalently $\textbf{A}=-\beta_E\nabla\alpha_E$, respectively, $\textbf{B}=\nabla\alpha_E\times\nabla\beta_E=\nabla\times\textbf{A}$ and therefore $\textbf{A}\cdot\textbf{B}=0$.

The gradients of $\alpha_E$ and $\beta_E$ can be expressed as
\begin{eqnarray}
\chi_i^{\mu\nu}(\nabla\alpha_E)_i^\mu&=&-\sum_j m_j(\alpha_{E,i}-\alpha_{E,j})(\nabla_i W_{ij}(h_i))^\nu,\\
\chi_i^{\mu\nu}(\nabla\beta_E)_i^\mu&=&-\sum_j m_j(\beta_{E,i}-\beta_{E,j})(\nabla_i W_{ij}(h_i))^\nu,
\end{eqnarray}
where
\begin{eqnarray}
  \chi_i^{\mu\nu}=\sum_j m_j(\textbf{r}_j-\textbf{r}_i)^\mu(\nabla_i W_{ij}(h_i))^\nu,
\end{eqnarray}
which is exact for linear functions (\citealp{Price&Bate2007}), i.e. for initial conditions with an uniform field.
However, the difference in using this exact-linear interpolation compared to the usual gradient operator is marginal (D. Price, private communication).

 Unfortunately,  not all magnetic field configurations can be expressed in terms of Euler potentials easily as they enter eq. \ref{euler} in a nonlinear way, and they are not unique for certain field configurations (\citealp{Stern1970}, \citealp{Yahalom&Lynden-Bell2006}). The former problem restricts only the choice of the initial magnetic field, whereas the latter can be crucial. If there are field configurations which can be expressed by different sets of Euler potentials, then this implies, that some other field configurations cannot be expressed at all using Euler potentials. However, since the fields considered in this work are topologically `simple', we do not expect to encounter these problems.

Furthermore, Euler potentials do not allow to follow the winding of magnetic fields beyond a certain point. This constraint is due to the fact that using the Euler potentials, the magnetic field is essentially mapped on the initial particle arrangement. If the initial arrangement evolves too much during the simulation, particles carrying conflicting values of Euler potentials (i.e. values, which do no longer allow for a finite and unambiguous calculation of their gradients) can come close. Then, the ability of the Euler potentials to represent the magnetic field correctly is lost. This conflict is expected to occur when the magnetic field is wound up more than once, which poses a problem especially towards the central region of a simulated galaxy.

\subsubsection{Timestepping}
In $\textsc{Vine}$, there are basically three different time step criteria, based on changes in the acceleration of a particle,
\begin{eqnarray}
  \Delta t_a^{n+1}=\tau_\mathrm{acc}\sqrt{\frac{\epsilon}{|\textbf{a}|}},\label{ts1}
\end{eqnarray}
its velocity,
\begin{eqnarray}
  \Delta t_v^{n+1}=\tau_\mathrm{acc}\frac{\epsilon}{|\textbf{v}|},\label{ts2}
\end{eqnarray}
or both in combination,
\begin{eqnarray}
  \Delta t_{va}^{n+1}=\tau_\mathrm{acc}\frac{|\textbf{v}|}{|\textbf{a}|},\label{ts3}
\end{eqnarray}
where $\epsilon$, $\textbf{a}$ and $\textbf{v}$ are the gravitational softening length, the acceleration and velocity of a particle in the previous time step ($n$), respectively, and $\tau_\mathrm{acc}$ is an accuracy parameter.\\
Two additional time step criteria are applied in SPH simulations:
First, the Courant-Friedrichs-Lewy criterion as suggested by \citet{Monaghan1989},
\begin{eqnarray}
  \Delta t_\mathrm{CFL}^{n+1}=\tau_\mathrm{CFL}\frac{h_i}{c_s+1.2(\alpha_i c_s + \beta_i h_i \mathrm{max}_j \mu_{ij})},\label{ts4}
\end{eqnarray}
where $\alpha_i$ and $\beta_i$ are artificial viscosity parameters, $c_s$ is the sound speed, $h_i$ the SPH softening length for gas particle $i$, and $\mu_{ij}$ corresponds to the velocity divergence between particles $i$ and $j$ with the maximum taken over all neighboring particles $j$ of particle $i$ (see \citealp{VINEI} for more details).
Secondly, there is a limit on how much the SPH softening lengths are allowed to change during one timestep:
\begin{eqnarray}
  \Delta t_h^{n+1}=\tau_h\frac{h_i}{\dot{h}_i},\label{ts5}
\end{eqnarray}
where $\tau_h$ is again an accuracy parameter. Usually, we apply $\tau_\mathrm{acc}=1$, $\tau_\mathrm{CFL}=0.5$ and $\tau_h=0.15$. The timestep actually employed in the simulation is the minimum of the timesteps in eqs. \ref{ts1}-\ref{ts5}.

\subsection{$\textsc{Gadget}$}

A somewhat different treatment of hydrodynamics and magnetic
fields is realised within the MPI parallel \textsl{N}-body/SPH code
$\textsc{Gadget}$ (\citealp{SpringelGadget1}, \citealp{SpringelGadget},
\citealp{GadgetMHD}). There are two significant differences in the
implementation relevant even for non-radiative simulations:

First, $\textsc{Vine}$ follows a classical implementation which is
integrating the internal energy, whereas $\textsc{Gadget}$ utilises what
is generally called the entropy conserving formulation. The important difference thereby is not
the fact that $\textsc{Gadget}$ integrates the entropy instead of the
internal energy. The crucial differences are rather the way in which the smoothing length $h_i$ is defined (in $\textsc{Gadget}$, $h_i$ is defined based on the mass within the kernel instead of the number of particles) and the inclusion of correction terms arising from the varying smoothing length. Also, the entropy conserving formulation uses a way of symmetrizing the kernel given by the derivation of the SPH equations, which in sum leads to conservation of energy and
entropy at the same time (\citealp{Springel&Hernquist2002}).

The second difference originates in an alternative formulation of the artificial
viscosity. In $\textsc{Gadget}$, artificial viscosity is based on the signal velocity instead of sound speed (\citealp{Monaghan1997}) and apt to incorporate magnetic waves in a natural way (\citealp{Price&Monaghan2004SPMHDI}).

This different implementation was shown to bring measurable improvements specially for MHD applications (\citealp{GadgetMHD}), but should not make too much of a difference for passive magnetic fields. The implementation of the induction equation and the Euler potentials formalism is the same in both codes.

The integration in $\textsc{Gadget}$ is also performed using
the leapfrog integration scheme, but $\textsc{Gadget}$ utilises a
kick-drift-kick-scheme whereas $\textsc{Vine}$ uses a
drift-kick-drift-scheme.

The timestep is given by
\begin{eqnarray}
  \Delta t^{n+1}=\sqrt{\frac{2\eta\epsilon}{|\textbf{a}|}},\label{ts6}
\end{eqnarray}
where $\eta$ translates to the accuracy parameter $\tau_\mathrm{acc}$ in eq. \ref{ts1} via $\tau_\mathrm{acc}=\sqrt{2\eta}$.
For SPH particles, also a Courant-like condition in the form
\begin{eqnarray}
  \Delta t^{n+1}_\mathrm{cour}=\frac{C_\mathrm{cour} h_i}{\mathrm{max}_jv_{ij}^\mathrm{sig}}\label{ts7}
\end{eqnarray}
is applied, where $h_i$ is the SPH softening length for gas particle $i$ and $v_{ij}^\mathrm{sig}$ the signal velocity between particles $i$ and $j$ as defined in \cite{Price&Monaghan2004SPMHDI} with the maximum taken over all neighboring particles $j$ of particle $i$. $C_\mathrm{cour}$ is an accuracy parameter which does not translate one-to-one to $\tau_\mathrm{CFL}$ in eq. \ref{ts4} due to the different definition of the Courant criterion.
We commonly use values of $\eta=0.02$ and $C_\mathrm{cour}=0.15$ to ensure that the timestep $\Delta t$ in $\textsc{Gadget}$ does not get too large compared to $\textsc{Vine}$. However, changing the accuracy parameters by a factor of two does not affect the overall evolution and amplification of the magnetic field in the simulated systems (not shown).

Beside that, the codes differ in details
of the tree construction for calculating gravitational forces.
For more details we refer the reader to the code papers for
$\textsc{Vine}$ (\citealp{VINEI}, \citealp{VINEII}) and
$\textsc{Gadget}$ (\citealp{SpringelGadget}, \citealp{GadgetMHD}).

\section{Simulations}
\label{SIMULATIONS}

\subsection{Setup}

The initial conditions for our Milky Way like galaxy are realised using the method described by \citet{Springel2005} which is based on \citet{Hernquist1993} (see also \citealp{Johansson2009}). The galaxy consists of an exponential stellar disc and a flat extended gas disc, a stellar bulge and a dark matter halo of collisionless particles. The gas is represented by SPH particles adopting an isothermal equation of state with a fixed sound speed of $c_s\approx 15$ km s$^{-1}$, which corresponds to a temperature of $T\approx2\cdot10^4$ K for a molecular weight of $1.4/1.1\cdot m_\mathrm{proton}$. We briefly note that by using an isothermal equation of state only one component of the ISM is modeled, typically this is a reasonably good approximation for the warm gas phase in disc galaxies (e.g. \citealp{Barnes2002}, \citealp{Li2005}, \citealp{Naab&Jesseit2006}). Assuming an isothermal equation of state implies that additional heat created in shocks by adiabatic compression and feedback processes (e.g. by SNII) is radiated away immediately. In addition, substantial heating processes prevent the gas from cooling below its effective temperature predefined by its sound speed.

The parameters describing the initial conditions can be found in Table \ref{tab1}. The particle numbers and the gravitational and SPH softening lengths used in the different runs can be found in Table \ref{tab2}.

\begin{table}
{\footnotesize \begin{tabular}{|l|lcr|}
  \hline
  total mass & $M_\mathrm{tot}$ & = &  $1.34\cdot 10^{12}M_\odot$\\
  disc mass & $M_\mathrm{disc}$ & = & $0.041\cdot M_\mathrm{tot}$ \\
  bulge mass & $M_\mathrm{bulge}$ & = & $0.01367\cdot M_\mathrm{tot}$ \\
  mass of the extended gas disc & $M_\mathrm{gas}$ & = & $0.2\cdot M_\mathrm{disc}$ \\
  exponential disc scale length & $l_D$ & = & $3.5$ kpc \\
  scale height of the disc & $h$ & = & $0.2\cdot l_D$ \\
  bulge scale length & $l_B$ & = & $0.2\cdot l_D$ \\
  extent of flat gas disc& $l_G$ & = & $6\cdot l_D$ \\
  spin parameter & $\lambda$ & =& 0.033\\
  virial velocity of the halo & $v_\mathrm{vir}$ & = & 160 km s$^{-1}$ \\
  half mass circular velocity & $v_\mathrm{half}$ & $\approx$ & 200 km s$^{-1}$ \\
  half mass rotation period & $T_\mathrm{half}$ & $\approx$ & 150 Myr \\
  isothermal sound speed & $c_s$ & $\approx$ & 15 km s$^{-1}$  \\
  initial magnetic field & $B_0$ & = & $10^{-9}$ G \\\hline
\end{tabular}}
  \caption{Parameters of initial disc setup}\label{tab1}
\end{table}

\begin{table*}
 { \begin{tabular}{|l|c|c|c|c|c|c|}\hline
   & \multicolumn{2}{|c|}{low resolution} & \multicolumn{2}{|c|}{normal resolution} & \multicolumn{2}{|c|}{high resolution}\\
 & \multicolumn{6}{|c|}{number of particles}\\\hline
  Halo & \multicolumn{2}{c}{$6\cdot10^4$} & \multicolumn{2}{c}{$6\cdot10^5$} & \multicolumn{2}{c}{$6\cdot10^6$} \\
  Disc & \multicolumn{2}{c}{$3\cdot10^4$} & \multicolumn{2}{c}{$3\cdot10^5$} & \multicolumn{2}{c}{$3\cdot10^6$} \\
  Bulge& \multicolumn{2}{c}{$1\cdot10^4$} & \multicolumn{2}{c}{$1\cdot10^5$} & \multicolumn{2}{c}{$1\cdot10^6$} \\
  Gas  & \multicolumn{2}{c}{$3\cdot10^4$} & \multicolumn{2}{c}{$3\cdot10^5$} & \multicolumn{2}{c}{$3\cdot10^6$}  \\
  Total& \multicolumn{2}{c}{$13\cdot10^4$} & \multicolumn{2}{c}{$13\cdot10^5$} & \multicolumn{2}{c}{$13\cdot10^6$} \\\hline
  & \multicolumn{6}{|c|}{fixed gravitational softening lengths $\epsilon$ [kpc]}\\\hline
  & $\textsc{Vine}$ & $\textsc{Gadget}$ & $\textsc{Vine}$ & $\textsc{Gadget}$ & $\textsc{Vine}$ & $\textsc{Gadget}$ \\\hline
  Halo  & 0.934/2 & 0.934 & 0.434/2 & 0.434 & - & 0.199 \\
  Disc  & 0.248/2 & 0.248 & 0.114/2 & 0.114 & - & 0.052 \\
  Bulge & 0.269/2 & 0.269 & 0.127/2 & 0.127 & - & 0.059 \\
  Gas   & 0.248/2 & 0.248 & 0.114/2 & 0.114 & - & 0.052  \\\hline
  & \multicolumn{6}{c}{minimum SPH softening lengths $h_\mathrm{min}$}\\\hline
  Gas   & $0.01\epsilon$  & $0.01\epsilon$ & $0.01\epsilon$ & $0.01\epsilon$ & - & $0.01\epsilon$\\\hline
\end{tabular}}
  \caption{Particles numbers and softening lengths. The factor two accounts for the different definition of the Kernel extent in $\textsc{Vine}$ ($\varrho < 2$) and $\textsc{Gadget}$ ($\varrho < 1$).}\label{tab2}
\end{table*}

Before we include magnetic fields we allow the galaxy to evolve for approximately three half mass rotation periods. For simplicity we choose an initial magnetic field in the $x$ direction. Its value, $B_0=10^{-9}$ G, corresponds to the typical value of intergalactic magnetic fields (\citealp{KronbergLesch&Hopp1999}). To set up the corresponding Euler potentials, we choose
\begin{eqnarray}
  \alpha_\mathrm{E}&=&B_0\cdot y,\\
  \beta_\mathrm{E}&=&y+z.
\end{eqnarray}

\begin{figure}
\begin{center}
  \epsfig{file=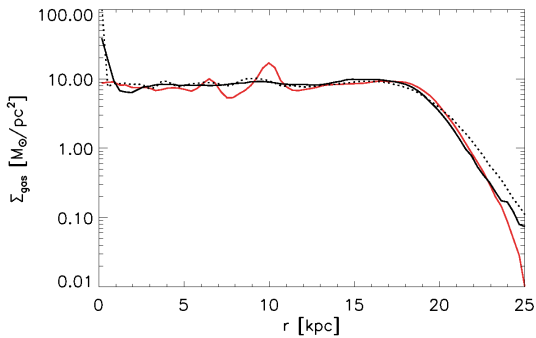, width=0.4\textwidth}
  \caption{Surface densities $\Sigma_\mathrm{gas}$ of the extended gas discs as a function of radius before the inclusion of the magnetic fields after 0.5 Gyr (red line) and after 2 Gyr (black lines) for simulations with $\textsc{Gadget}$ (solid line) and $\textsc{Vine}$ (dotted line). The gas discs are stable for more than ten half mass rotation periods.}
  {\label{sigma_gas}}
\end{center}
\end{figure}
\begin{figure}
\begin{center}
  \epsfig{file=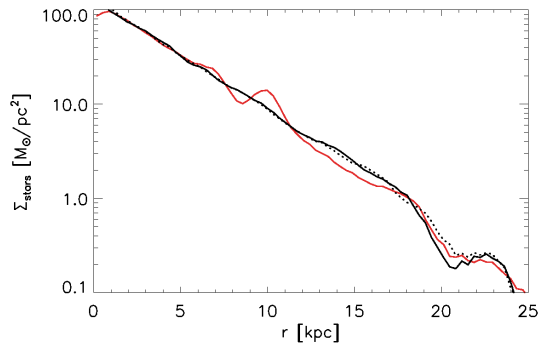, width=0.4\textwidth}
  \caption{Same as Fig. \ref{sigma_gas} but for the stellar surface densities $\Sigma_\mathrm{stars}$. Both the stellar and the gas discs are stable for more than ten half mass rotation periods.}
  {\label{sigma_disc}}
\end{center}
\end{figure}
\begin{figure}
\begin{center}
  \epsfig{file=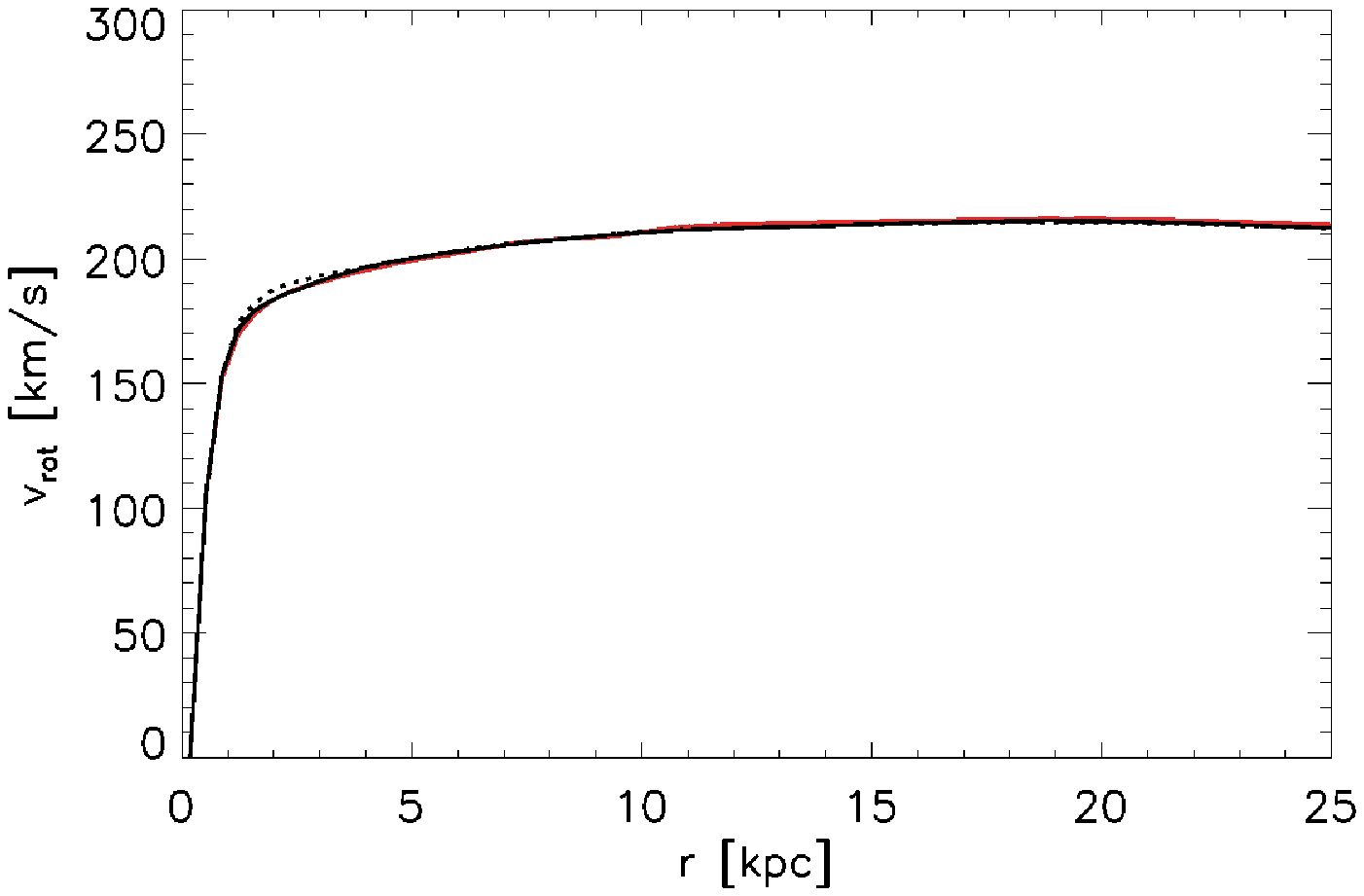, width=0.4\textwidth}
  \caption{Circular velocity curves of the simulated galaxies at two different times. The colour coding is the same as in Figs. \ref{sigma_gas} and \ref{sigma_disc}. Again, the circular velocity curves are stable over more than ten half mass rotation periods.}
  {\label{vrot}}
\end{center}
\end{figure}

We have checked the stability of our discs in independent simulations without magnetic fields.
Figs. \ref{sigma_gas} and \ref{sigma_disc} show the surface densities $\Sigma_\mathrm{gas}$ of the extended gaseous discs and $\Sigma_\mathrm{stars}$  of the exponential stellar discs, respectively, as a function of radius for $t=0.5$ Gyr (red), i.e. the time at which the magnetic field is switched on, and $t=2.0$ Gyr (black). Fig. \ref{vrot} shows the circular velocity curves of the simulated galaxies at the same times. The discs simulated with $\textsc{Vine}$ (dotted line) and $\textsc{Gadget}$ (solid line) show similar results and stay stable over more than ten half mass rotation periods.

\subsection{Direct magnetic field simulations}

\begin{figure*}
\centering
  \epsfig{file=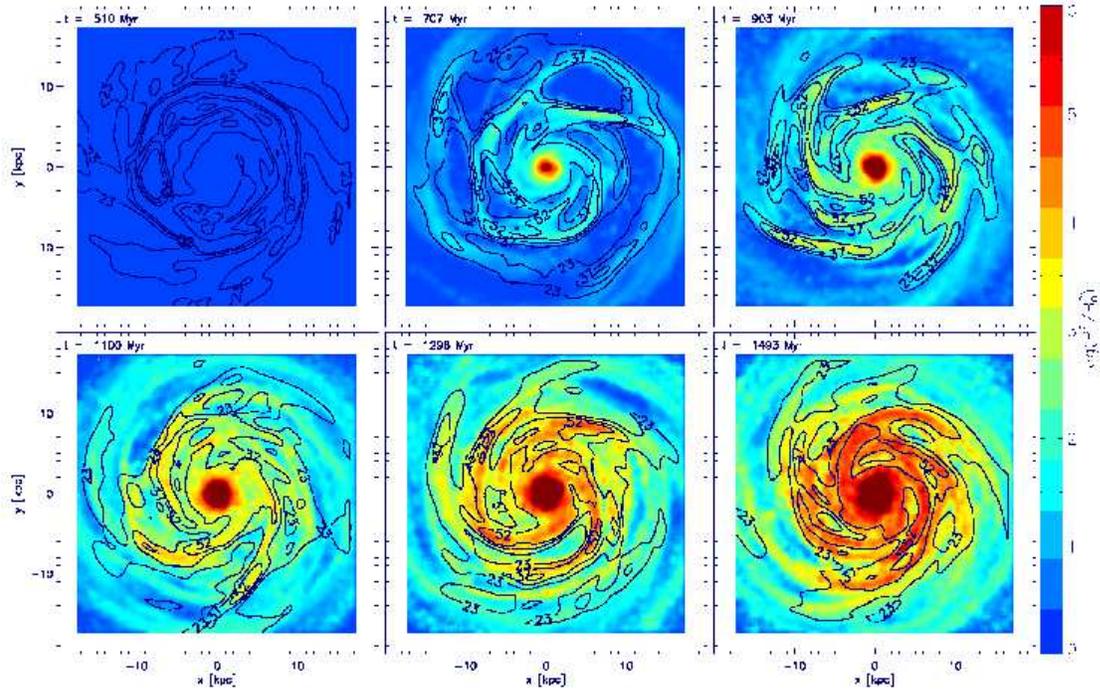, width=0.84\textwidth}
  \caption{Face-on magnetic field energy and gas density as a function of time for the simulation performed with $\textsc{Vine}$ using direct magnetic field description and without applying the viscosity limiter. The colours correspond to the magnetic field energy $B^2/8\pi$ on a logarithmic scale, normalised to the initial value of $\frac{1}{8\pi}\cdot 10^{-18}$ erg cm$^{-3}$. The contour lines indicate physical densities of $23$, $37$ and $52$ $M_\odot$ pc$^{-3}$, respectively.}
  {\label{indVINE}}
\end{figure*}

\begin{figure*}
\centering
  \epsfig{file=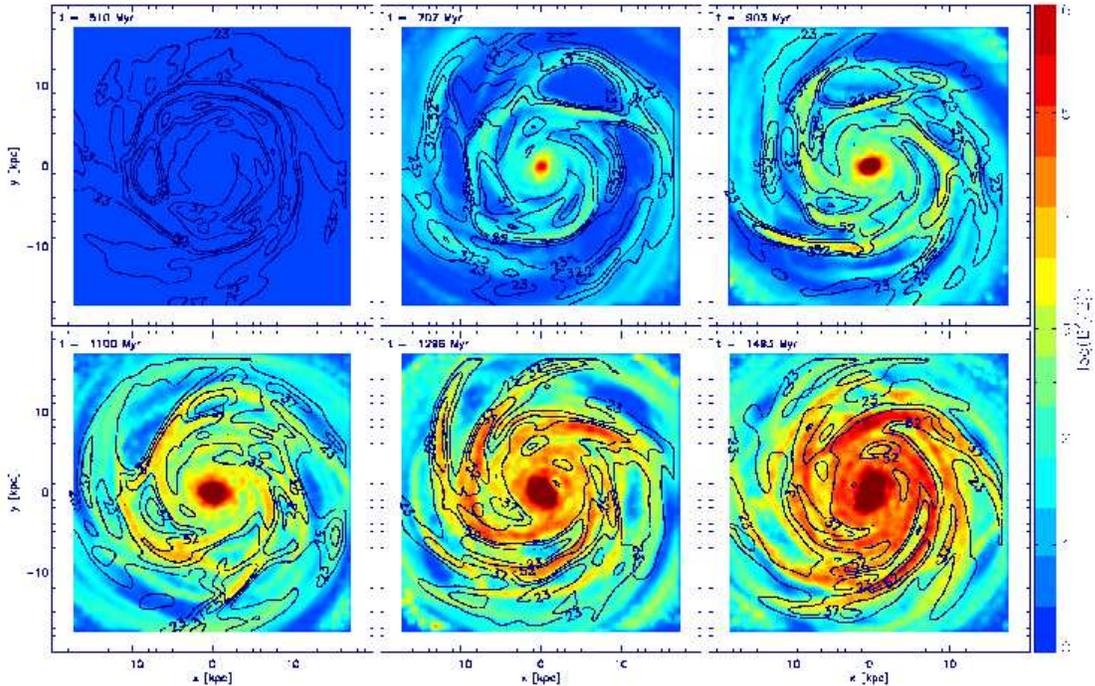, width=0.84\textwidth}
  \caption{Same as Fig. \ref{indVINE} for identical initial conditions simulated with $\textsc{Gadget}$. The morphology is very similar but the magnetic field reaches higher values in the spiral arms in the $\textsc{Gadget}$ simulation compared to the simulation with $\textsc{Vine}$.}
  {\label{indGAD}}
\end{figure*}

Figs. \ref{indVINE} and \ref{indGAD} show the face on view of the magnetic field energy and gas density of the simulated galaxy at different output times. The magnetic field was switched on at $t=510$ Myr. The viscosity limiter was not applied. Fig. \ref{indVINE} shows the simulation performed with $\textsc{Vine}$ and Fig. \ref{indGAD} the same initial conditions simulated with $\textsc{Gadget}$. The magnetic field energy $B^2/8\pi$ is colour coded and normalised to the initial value of $\frac{1}{8\pi}\cdot 10^{-18}$ erg cm$^{-3}$ on a logarithmic scale from 1 (blue) to $1.5\cdot 10^{8}$ (red). The contours overplotted indicate physical densities of $23$, $37$ and $52$ $M_\odot$ pc$^{-3}$, respectively. We use a grid with a cell size of 0.3 kpc for the calculation of the mean values of the densities and the magnetic field energies, averaging in the vertical direction from -$h$ to $h$, where $h$ is the local height of the gas disc.

\begin{figure}
\centering
  \epsfig{file=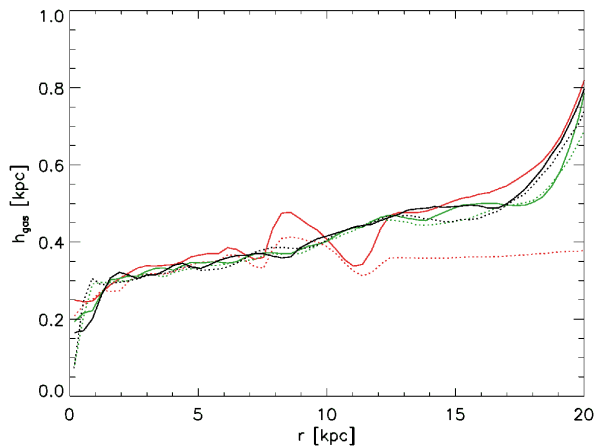, width=0.45\textwidth}
  \caption{SPH softening lengths $h_\mathrm{gas}$ as a function of radius shortly after the inclusion of the magnetic fields at 0.55 Gyr (red line), at 1.25 Gyr (green lines) and at 2 Gyr (black lines) for simulations with $\textsc{Gadget}$ (solid lines) and $\textsc{Vine}$ (dotted lines). The SPH smoothing lengths are very similar for both codes.}
  {\label{hsml}}
\end{figure}

 In both simulations we see that the magnetic field energy pattern is tightly connected to the density pattern of the gas. Moreover, both simulated galaxies show a very similar morphology in the gas and magnetic field distributions. The magnetic field energy in the spiral arms is amplified by up to five orders of magnitude in both codes and even more in the central region (see also Fig. \ref{Bwithr}). Furthermore, the SPH smoothing lengths $h_\mathrm{gas}$ are similar for both codes (Fig. \ref{hsml}), indicating that the performance of the hydrodynamic calculations is concerted. The smoothing lengths in $\textsc{Vine}$ are initially set to a constant value of $h_\mathrm{gas}\approx 0.3$ kpc at the time of the magnetic field inclusion.

\begin{figure*}
\centering
  \epsfig{file=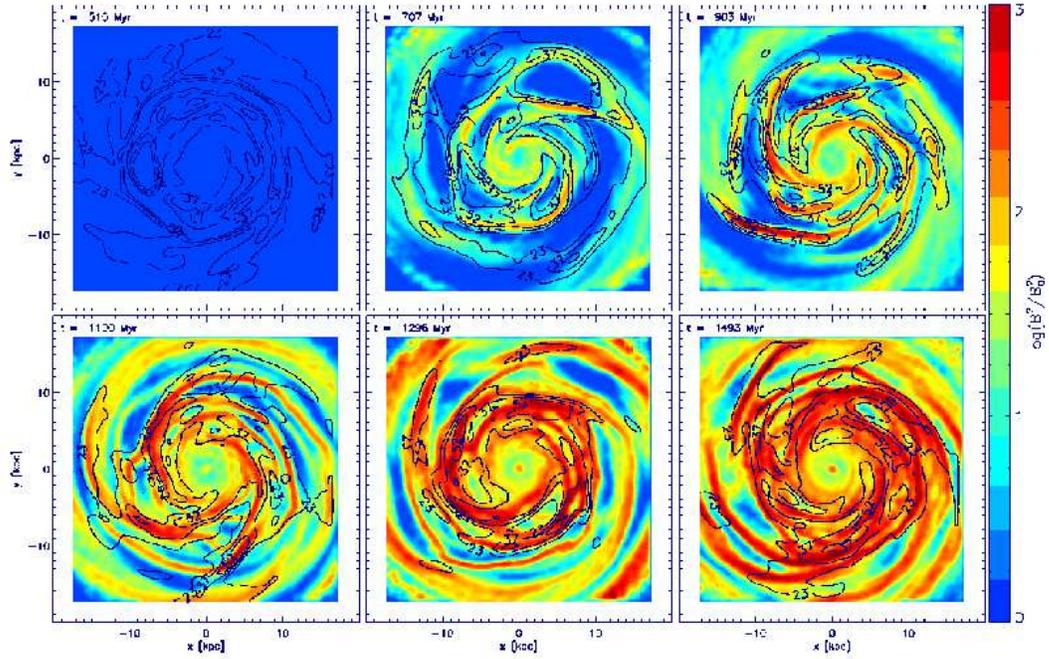, width=0.8\textwidth}
  \caption{Same as Fig. \ref{indVINE}, this time the magnetic field is followed using Euler potentials implemented in $\textsc{Vine}$. In contrast to the direct simulation the magnetic field is more strongly amplified in the spiral arms than at the centre. The maximum amplification of the magnetic field energy is only three orders of magnitude. Note that the colour scaling is different to Figs. \ref{indVINE} and \ref{indGAD}.}
  {\label{eulerVINE}}
\end{figure*}

\begin{figure*}
\centering
  \epsfig{file=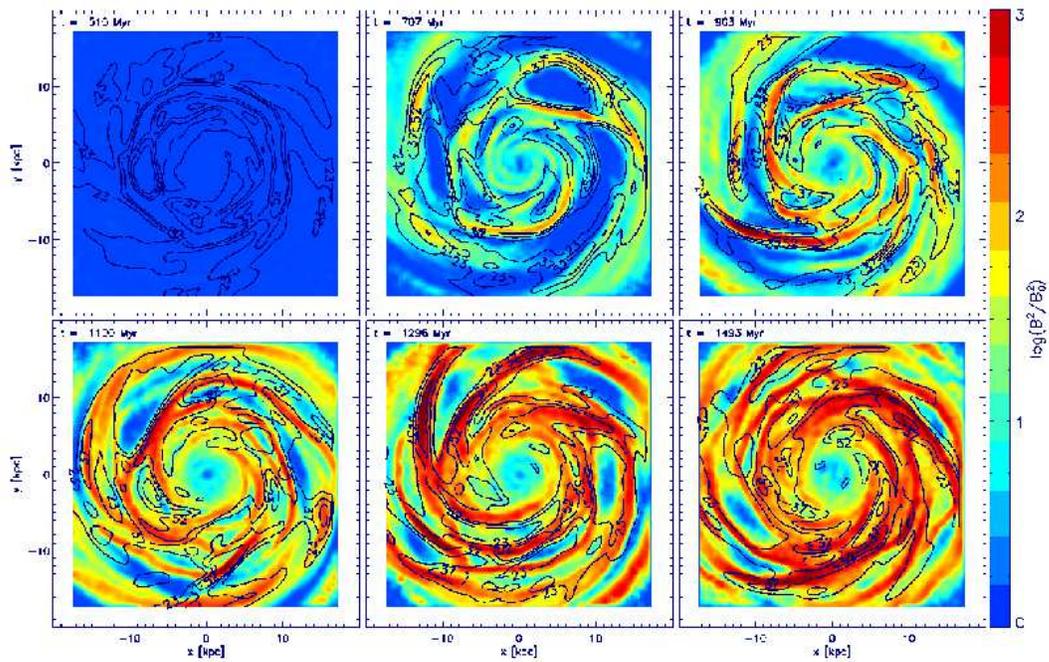, width=0.8\textwidth}
  \caption{Same as Fig. \ref{indGAD}, this time the magnetic field is followed using Euler potentials implemented in $\textsc{Gadget}$. The energies and morphology of the magnetic field is now similar to the $\textsc{Vine}$ simulation.}
  {\label{eulerGAD}}
\end{figure*}

\subsection{SPH with Euler Potentials}

Figs. \ref{eulerVINE} and \ref{eulerGAD} show simulations starting from the same initial conditions as before. However, this time the evolution of the magnetic field was followed using the Euler potentials. Again, we show  magnetic field energies and gas densities. This time the amplification of the magnetic field energy in the spiral arms is only three orders of magnitude for both simulations with $\textsc{Vine}$ and $\textsc{Gadget}$, with both showing a remarkably similar evolution. The most notable difference to the simulations with direct magnetic field treatment shown in Fig. \ref{indVINE} and \ref{indGAD} is at the centre of the galaxies, where in the direct simulations the field amplification was strongest. With Euler potentials the magnetic field grows mostly in the spiral arms of the galaxy (see also Fig. \ref{Bwithr}).

\begin{figure}
\begin{center}
  \epsfig{file=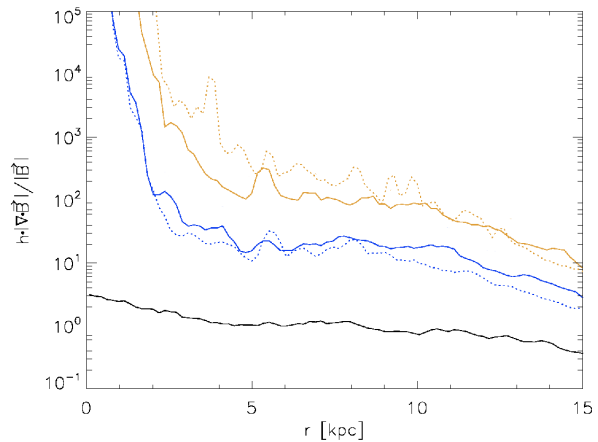, width=0.45\textwidth}
  \caption{Numerical $h\cdot|\nabla\cdot \textbf{B}|/|\textbf{B}|$ at $t\approx 1.5$ Gyr as a function of radius for simulations without applying the viscosity limiter using direct magnetic field treatment (blue) and using Euler potentials (black) in $\textsc{Gadget}$ (solid lines) and $\textsc{Vine}$ (dotted lines). Direct magnetic field simulations for which the viscosity limiter was applied are also shown (orange). Using direct magnetic field description, the numerical $h\cdot\nabla\cdot \textbf{B}$ is highest at small radii, and much larger than in the Euler potentials formalism.}
  {\label{divergence_radius}}
\end{center}
\end{figure}

Since the magnetic fields in our simulations are passive, the density profiles (Figs. \ref{sigma_gas} and \ref{sigma_disc}) of the disc are the same for all runs. Thus, the different profiles of the magnetic field energy cannot be traced back to the density profiles.
In fact, it is the numerical $\nabla\cdot \textbf{B}$ which presumably causes the high amplification of the magnetic field at the centre in simulations with the direct magnetic field treatment. Fig. \ref{divergence_radius} shows the radial profile of the numerical $h\cdot|\nabla\cdot \textbf{B}|/|\textbf{B}|$ at time $t\approx 1.5$ Gyr for simulations using direct magnetic field treatment (blue for simulations without applying the viscosity limiter and orange where the limiter was applied) and Euler potentials (black) performed using $\textsc{Gadget}$ (solid lines) and $\textsc{Vine}$ (dotted line). Utilising the direct magnetic field description, the numerical $\nabla\cdot \textbf{B}$ is highest at small radii, and much larger than for the Euler potential formalism. As will be discussed in the following section, high $\nabla\cdot \textbf{B}$ corresponds to high amplification of the magnetic field.

\begin{figure}
\begin{center}
  \epsfig{file=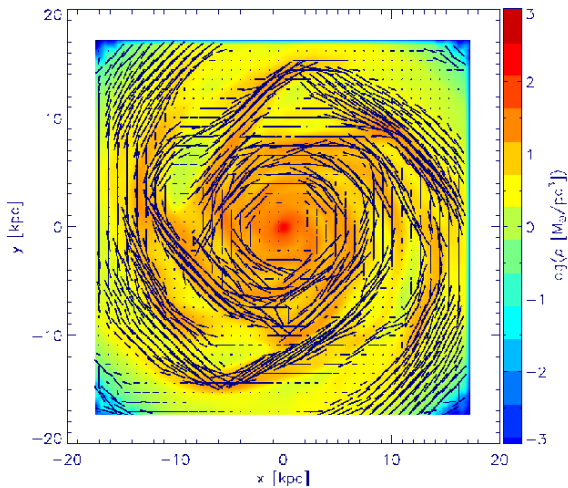, width=0.45\textwidth}
  \caption{Gas density (colour coded) and magnetic field vectors for the normal resolution simulation with $\textsc{Vine}$ using the Euler potentials formalism at $t\approx 1$ Gyr, i.e. $\approx$ 500 Myr after the inclusion of the magnetic field. The length of the vectors is normalised to the initial value and displayed logarithmically. $l=0$ corresponds to $ B\approx B_0$ or smaller and $l=3$ to $B=10\cdot B_0$.}
  {\label{vectors}}
\end{center}
\end{figure}

Fig. \ref{vectors} shows the magnetic field vectors for the normal resolution $\textsc{Vine}$ simulation utilising Euler potentials at the time $t\approx 0.9$ Gyr.  This time the colours correspond to the gas density on a logarithmic scale from $0.3\cdot 10^{-3}$ to $2.3\cdot 10^3 M_\odot$ pc$^{-3}$, overplotted with the field vectors. The length $l$ of the vectors is normalised to the initial value and displayed logarithmically as $l=3\cdot \log(B/B_0)$, i.e. $l=0$ corresponds to $ B\approx B_0$ or smaller, $l=1$ to $B\approx 2\cdot B_0$, $l=2$ to $B\approx 5\cdot B_0$ and $l=3$ to $B=10\cdot B_0$. The magnetic field lines follow the spiral structure of the gas. They have been amplified by contraction in regions of higher density and restructured by differential rotation of the galaxy. Their orientation is caused by the motion of the gas. These characteristics are very similar to typical observations of magnetic fields in galactic discs (e.g. Fig. \ref{M51}).

Qualitatively, this behaviour is the same for all simulations using both codes. Only the central region in simulations using direct magnetic field treatment shows chaotic orientation of the magnetic field lines, indicating artificial amplification of the magnetic field due to high numerical $\nabla\cdot \textbf{B}$.

\begin{figure*}
\centering
  \epsfig{file=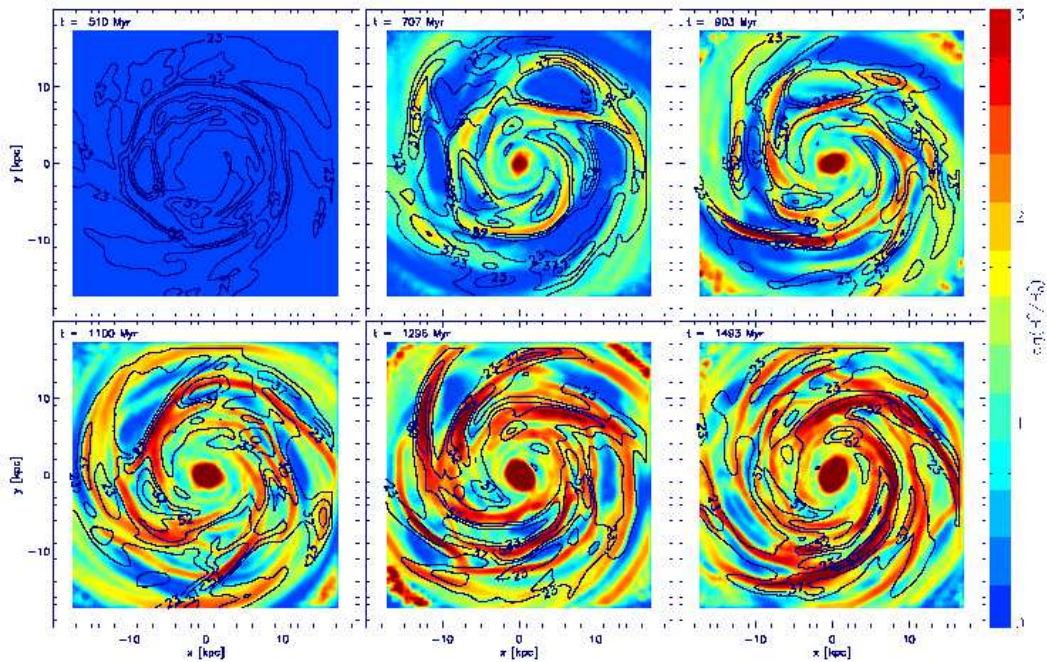, width=0.8\textwidth}
  \caption{Same as Fig. \ref{indGAD}, this time the magnetic field is smoothed every 30 timesteps. The field morphology is similar to the morphology in Fig. \ref{indGAD} and \ref{indVINE}, respectively, but the magnetic field values in the spiral arms are now more similar to the values in the Euler implementation.}
  {\label{smoothGAD}}
\end{figure*}

\section{Evaluation}
\label{DISCUSSION}

\subsection{Magnetic field growth}

Figs. \ref{indVINE} (\ref{indGAD}) and \ref{eulerVINE} (\ref{eulerGAD}), respectively, reveal the differences in the magnetic field amplification for the direct magnetic field treatment and the Euler potentials formalism: Using the direct description, the amplification of the magnetic field energy in the spiral arms is higher by at least two orders of magnitude, and at the centre even more than six orders of magnitude compared to the Euler potentials method. This difference is probably caused by the numerical $\nabla\cdot \textbf{B}$ in these simulations (Fig. \ref{divergence_radius}), but possibly also by the fact that field winding is not traced beyond a certain evolutionary state in the Euler potentials formulation (see section \ref{EULER}). Since the Euler potentials are free from physical divergence by construction (i.e. the divergence is zero to measurements errors), the numerical divergence in simulations using the Euler potentials is due to the SPH derivative approximation when calculating the magnetic field from the potentials (Eq. \ref{euler}). In this sense, the numerical divergence found in simulations using Euler potentials reflects the ability of SPH operators to measure the gradient of a curl to zero. Thus, the fact that $\nabla\cdot \textbf{B}$ is higher by approximately one order of magnitude in the disc (i.e. within $\approx 5$ to 15 kpc) and by several orders of magnitude at the centre (Fig. \ref{divergence_radius}), presumably causes the different magnetic field amplification in these simulations. This is the case at least in the disc region, where the winding of the field is not strong enough to constrain the Euler potentials formulation.

To get a better idea of the influence of numerical $\nabla\cdot \textbf{B}$ on the amplification of the magnetic filed, we have performed simulations applying magnetic field smoothing, a technique allowing for reduction of small scale fluctuations and therefore also the numerical divergence (\citealp{GadgetMHD}). Within this method, the magnetic field is smoothed periodically as suggested by \citet{Borve2001}. Fig. \ref{smoothGAD} shows again the magnetic field energies and gas densities for a $\textsc{Gadget}$ simulation starting from the same initial conditions as before and without applying the viscosity limiter. This time, the magnetic field was smoothed every 30 timesteps. Applying the smoothing scheme, the amplification of the magnetic field energy is reduced to approximately three orders of magnitude within the spiral arms, which is the same as the amplification seen in simulations using the Euler potentials, and it is also lowered towards the centre of the galaxy. The structure of the magnetic field is despite the smoothing still very similar to the other runs and again correlates well with the structure of the gas density, however, the magnetic field energy is more concentrated within the spiral arms.

\begin{figure}
\centering
  \epsfig{file=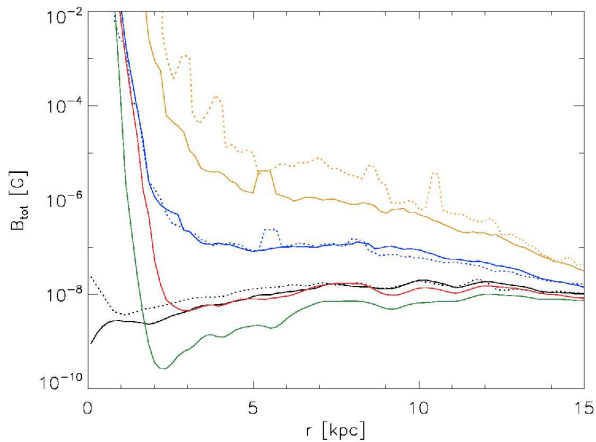, width=0.45\textwidth}
  \caption{Total magnetic field ($B_\mathrm{tot}=\sqrt{B_x^2+B_y^2+B_z^2}$) at $t\approx 1.5$ Gyr as a function of radius for the normal resolution $\textsc{Gadget}$ (solid lines) and $\textsc{Vine}$ (dotted lines) simulations without applying the viscosity limiter using Euler potentials (black) and the direct magnetic field description (blue). Direct magnetic field simulations for which the viscosity limiter was applied are also shown (orange). Two direct magnetic field implementations including field smoothing to reduce the numerical $\nabla \cdot \textbf{B}$ contribution are shown in red and green. The simulations including smoothing have been run with a smoothing interval of 30 (red) and 5 timesteps (green), respectively. Increasing the frequency of smoothing tends to decrease the amplitude of the magnetic field between 3 and 10 kpc, but has relatively little effect for larger radii.}
  {\label{Bwithr}}
\end{figure}

Fig. \ref{Bwithr} shows the total magnetic field at $t\approx 1.5$ Gyr as a function of radius for the normal resolution $\textsc{Gadget}$ (solid line) and $\textsc{Vine}$ (dotted line) simulations using Euler potentials (black) and the direct magnetic field description without applying the viscosity limiter (blue) and with the limiter turned on (orange), respectively. The direct magnetic field simulations including field smoothing are shown in red and green. They have been performed with a smoothing interval of 30 (red) and 5 timesteps (green), respectively, and without applying the viscosity limiter. As discussed before, the most notable difference between simulations with direct magnetic field treatment and the Euler implementation is at the centre of the galaxies. There, the amplification in the direct simulations is much stronger than in the Euler simulations. This behaviour could be at least partly physical, as there are high radial velocities and strong in- and outflows of gas in the central region (fig. \ref{VelandB}), resulting according to Eqs. \ref{induktionsglg_d} and \ref{induktionsglg_e} in an amplification of the magnetic field. In addition, also the azimuthal derivatives of the radial and toroidal velocity components are large at the very centre, which also could account for the violent amplification (see section \ref{INSPECTION}). On the other hand, the second term of eq. \ref{induktionsglg_e} does not play an important role, since $dv_\varphi/dr$ is large and therefore $d\Omega/dr$ small in the central region (by reason of solid body rotation). However, the high $\nabla \cdot\textbf{B}$ values at the centre make it difficult to distinguish between physical growth and numerical errors. Since the Euler potentials are also unreliable in this region (see section \ref{EULER}), it is not easy to decide which formalism is the most capable in describing the physics in the centre of the galaxy correctly. This is also true for the simulations including smoothing. Increasing the frequency of smoothing tends to decrease the amplitude of the magnetic field between 3 and 10 kpc, but has relatively little effect for larger radii. Interestingly, the large increase of $\textbf{B}$ in the centre is never smoothed away, which could indicate, that this behaviour is actually partly physical. For the simulation which applies smoothing every 30 timesteps (red), the amplification of the field at $r>3$ kpc is similar to the simulations with Euler potentials. Applying smoothing every 5 timesteps (green), the amplification is considerably weaker than in the Euler potentials simulations, indicating that by such strong smoothing essential physics is lost, in agreement with earlier findings by \cite{GadgetMHD}.

In the following, we only consider the disc region (from 5 to 15 kpc), since the high numerical divergence in the centre makes it difficult to lower it to the value of the divergence seen in simulations with Euler potentials (i.e. $h\cdot|\nabla\cdot\mathbf{B}|/|\mathbf{B}|\approx 1$), without smoothing the magnetic field structure too much.

\begin{figure}
\begin{center}
  \epsfig{file=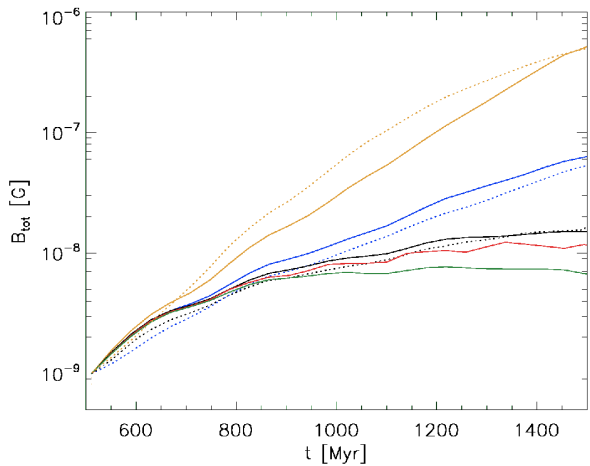, width=0.45\textwidth}
  \caption{Total magnetic field ($B_\mathrm{tot}=\sqrt{B_x^2+B_y^2+B_z^2}$) within the disc (between 5 and 15 kpc) as a function of time for different implementations. The colour coding is the same as in Fig. \ref{Bwithr}. Applying the smoothing scheme reduces the amplification of the magnetic field.}
  {\label{Bwithtime}}
\end{center}
\end{figure}

Fig. \ref{Bwithtime} shows the evolution of the total magnetic field ($B_\mathrm{tot}=\sqrt{B_x^2+B_y^2+B_z^2}$) within the disc with time for the different implementations. The colour coding is the same as in Fig. \ref{Bwithr}. As before, for the simulation which applies smoothing every 30 timesteps (red), the amplification of the field is similar to the simulation with Euler potentials. However, the performance of these simulation is not very convincing due to the ``jumps'' in the evolution caused by the artificial periodic smoothing. Applying smoothing every 5 timesteps (green), the amplification is as discussed before lower than in the Euler potentials simulations.

\begin{figure}
\begin{center}
  \epsfig{file=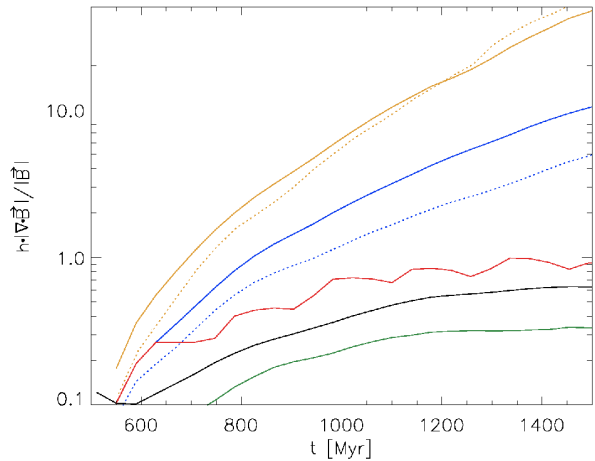, width=0.45\textwidth}
  \caption{Total divergence of the magnetic field within the disc (between 5 and 15 kpc) as a function of time for different implementations. The colour coding is the same as in Fig. \ref{Bwithr}. The higher the divergence, the stronger the amplification of the magnetic field.}
  {\label{DIVBwithtime}}
\end{center}
\end{figure}

This behaviour can be understood by considering the corresponding numerical divergence of the magnetic field. Fig. \ref{DIVBwithtime} shows $h\cdot|\nabla\cdot\textbf{B}|/|\textbf{B}|$ as a function of time for all simulations. In all cases, the growth of $h\cdot|\nabla\cdot\textbf{B}|/|\textbf{B}|$ behaves similar to the amplification of the total magnetic field, i.e. the higher the divergence, the stronger the amplification of the field. Though the numerical divergence in the simulation using Euler potentials (black) is higher than in the simulation with a smoothing interval of 5 timesteps (green), its value does not directly correlate with the field growth. That is because the (defective) magnetic field itself is not used for calculating the magnetic field evolution within the Euler potential formalism as is the case for the direct magnetic field description (compare eqs. \ref{IndEqsSPH} and \ref{euler}). Using the smoothing scheme lowers the divergence (in case of smoothing every 5 timesteps even below the numerical divergence of the Euler potential formalism) and lowers also the field amplification, leading (if applied not too often) to an amplification of the total field much more similar to that using the Euler potentials, which are free from \textsl{physical} divergence by construction.

Interestingly, for simulations applying the viscosity limiter suggested by \cite{Balsara1995}, the magnetic field amplification using the direct magnetic field description is in both codes much higher than without applying this limiter (orange lines in Figs. \ref{Bwithr} and \ref{Bwithtime}). The reason for this higher amplification is the higher velocity dispersion in these simulations. The viscosity limiter lowers the viscosity in regions of strong shear flows, thus suppressing velocity diffusion and leading to higher velocity gradients. Consistently, also the numerical divergence of the magnetic field is higher (and considerably higher than the ``unavoidable'' value of approximately one) in these simulations (orange lines in Figs. \ref{divergence_radius} and \ref{DIVBwithtime}). Applying the viscosity limiter in simulations using Euler potentials, however, does not change the evolution of the magnetic field significantly (not shown). Therefore, again, it is probable that the higher numerical $\nabla\cdot \mathbf{B}$ terms lead via the induction equation (eq. \ref{induktionsglg}) to an enhanced magnetic field growth.

In summary, the field amplification in case of direct magnetic field description correlates with the non-vanishing, numerical $\nabla\cdot\textbf{B}$. The Euler potential formalism also has its shortcomings (like the non-uniqueness and the dependence of the magnetic field on two derivatives (Eq. \ref{euler}) leading to lower numerical accuracy). Thus there is a strong need for simulations with different $\nabla\cdot \textbf{B}$ cleaning techniques and even higher resolution in order to be able to distinguish the best description for simulations of magnetic fields in galactic discs.

However, since the physical divergence is zero in the case of the Euler potentials, we believe this method (for the time being) to be the best one for our studies of magnetic convection in disc galaxies. The following discussion therefore concentrates on simulations using Euler potentials.

\subsection{Numerical resolution}

\begin{figure}
\begin{center}
  \epsfig{file=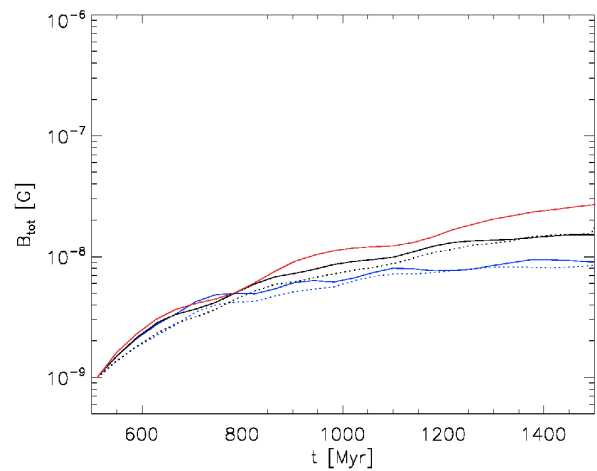, width=0.45\textwidth}
  \caption{Resolution study: Total magnetic field ($B_\mathrm{tot}=\sqrt{B_x^2+B_y^2+B_z^2}$) as a function of time for the $\textsc{Gadget}$ (solid line) and $\textsc{Vine}$ (dotted line) simulations without applying the viscosity limiter using Euler potentials. The total numbers of particles are $1.3\cdot10^5$ (blue), $1.3\cdot10^6$ (black) and $1.3\cdot10^7$ (red).}
  {\label{Bwithtime3res}}
\end{center}
\end{figure}

Fig. \ref{Bwithtime3res} shows the total magnetic field as a function of time for different resolutions (see Table \ref{tab2}) in simulations with $\textsc{Gadget}$ (solid lines) and $\textsc{Vine}$ (dashed lines) without applying the viscosity limiter.

One Gyr after its initialization the magnetic field has been amplified from $10^{-9}$ to approximately $9\cdot 10^{-9}$ G in the low resolution simulation (blue), whereas the final magnetic field strength in the normal resolution simulation is slightly more than 1.5 times higher ($1.5\cdot10^{-8}$ G). The final magnetic field strength in the high resolution run is again approximately 1.5 times higher than in the normal resolution run (i.e. $\approx 2.5\cdot10^{-8}$). The numerical $h\cdot|\nabla\cdot \textbf{B}|/|\textbf{B}|$ values are of the same order for all resolutions (not shown). Thus, we have not yet reached numerical convergence in the magnetic field evolution.

\subsection{Inspection of the induction equation}
\label{INSPECTION}

By analyzing the velocity and magnetic field in our simulation we can identify the single terms of the induction equation responsible for the behaviour of the magnetic field. Dropping all dependencies on $z$, the equations for the evolution of the radial and toroidal magnetic fields read
\\\\
\begin{tabular}{lcccc}
  $\displaystyle \frac{\partial B_r}{\partial t}$ & = &
  $\displaystyle -B_r\frac{v_r}{r}$ &
  $\displaystyle -\frac{1}{r}B_r\frac{\partial v_\varphi}{\partial\varphi}$ &
  $\displaystyle +\frac{1}{r}B_\varphi\frac{\partial v_r}{\partial\varphi}$ \\
  & & {\footnotesize (1)} & {\footnotesize (2)} & {\footnotesize (3)} \\[1.5ex]
  & & $\displaystyle -v_r\frac{\partial B_r}{\partial r}$ &
  $\displaystyle -\frac{v_\varphi}{r}\frac{\partial B_r}{\partial\varphi}$,\\
  & & {\footnotesize (4)} & {\footnotesize (5)} &  \\[1.5ex]
  $\displaystyle \frac{\partial B_\varphi}{\partial t}$ & = &
  $\displaystyle -B_\varphi\frac{\partial v_r}{\partial r}$ &
  $\displaystyle +B_r\frac{\partial v_\varphi}{\partial r}$ &
  $\displaystyle -v_r\frac{\partial B_\varphi}{\partial r}$\\
  & & {\footnotesize (6)} & {\footnotesize (7)} & {\footnotesize (8)} \\[1.5ex]
  & & $\displaystyle -\frac{v_\varphi}{r}\frac{\partial B_\varphi}{\partial\varphi}$ &
  $\displaystyle -\frac{v_\varphi}{r}B_r$,\\
  & & {\footnotesize (9)} & {\footnotesize (10)} & \\
\end{tabular}\\\\
where we have labelled the single terms with numbers for easier reference.

\begin{figure*}
\begin{center}
  \epsfig{file=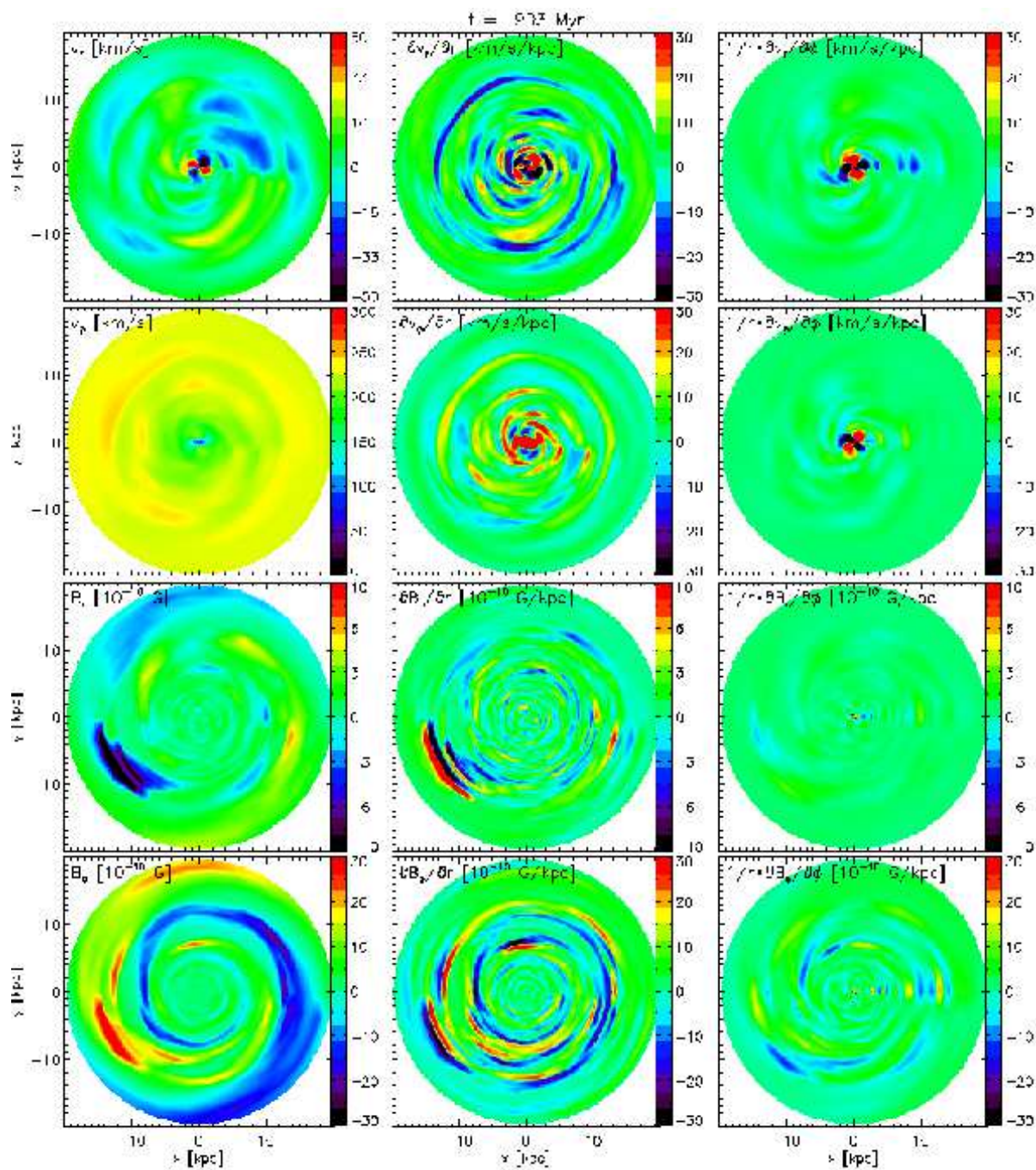, width=0.9\textwidth}
  \caption{Radial and toroidal components of the velocity and the magnetic field and their derivatives at $t \approx 900$ Myr for the normal resolution $\textsc{Gadget}$ simulation. From left to right and top to bottom:
  $v_r$,
  $\frac{\partial v_r}{\partial r}$,
  $\frac{1}{r}\cdot\frac{\partial v_r}{\partial \varphi}$,
  $v_\varphi$,
  $\frac{\partial v_\varphi}{\partial r}$,
  $\frac{1}{r}\cdot\frac{\partial v_\varphi}{\partial \varphi}$,
  $B_r$,
  $\frac{\partial B_r}{\partial r}$,
  $\frac{1}{r}\cdot\frac{\partial B_r}{\partial \varphi}$,
  $B_\varphi$,
  $\frac{\partial B_\varphi}{\partial r}$
   and $\frac{1}{r}\cdot\frac{\partial B_\varphi}{\partial \varphi}$.}
  {\label{VelandB}}
\end{center}
\end{figure*}

The radial and toroidal components of the velocity and the magnetic field and their corresponding derivatives are shown in Fig. \ref{VelandB} after approximately three half mass rotation periods after the onset of the magnetic field. The radial velocity (top left) is typically negative, leading to an effective gas inflow towards the centre of the galaxy. This negative radial velocity mirrors the angular momentum transport to large radii of the galaxy by spiral arm formation. The mean circular velocity (second row, left) is 210 km s$^{-1}$ at large radii, and drops to zero towards the centre (see also Fig. \ref{vrot}). The toroidal magnetic field (bottom left) is wound up by differential rotation, leading to a structure of altering positive and negative magnetic field values from centre to the edge of the galaxy. Consequently the derivatives with respect to $\varphi$ (right panel) are smaller than the radial derivatives (middle panel), mirroring the approximate axial symmetry. However, since the terms of the induction equation depend always on a product between a derivative and a velocity or magnetic field component, one cannot a priori neglect the terms depending on azimuthal derivatives.

\begin{figure}
\begin{center}
  \epsfig{file=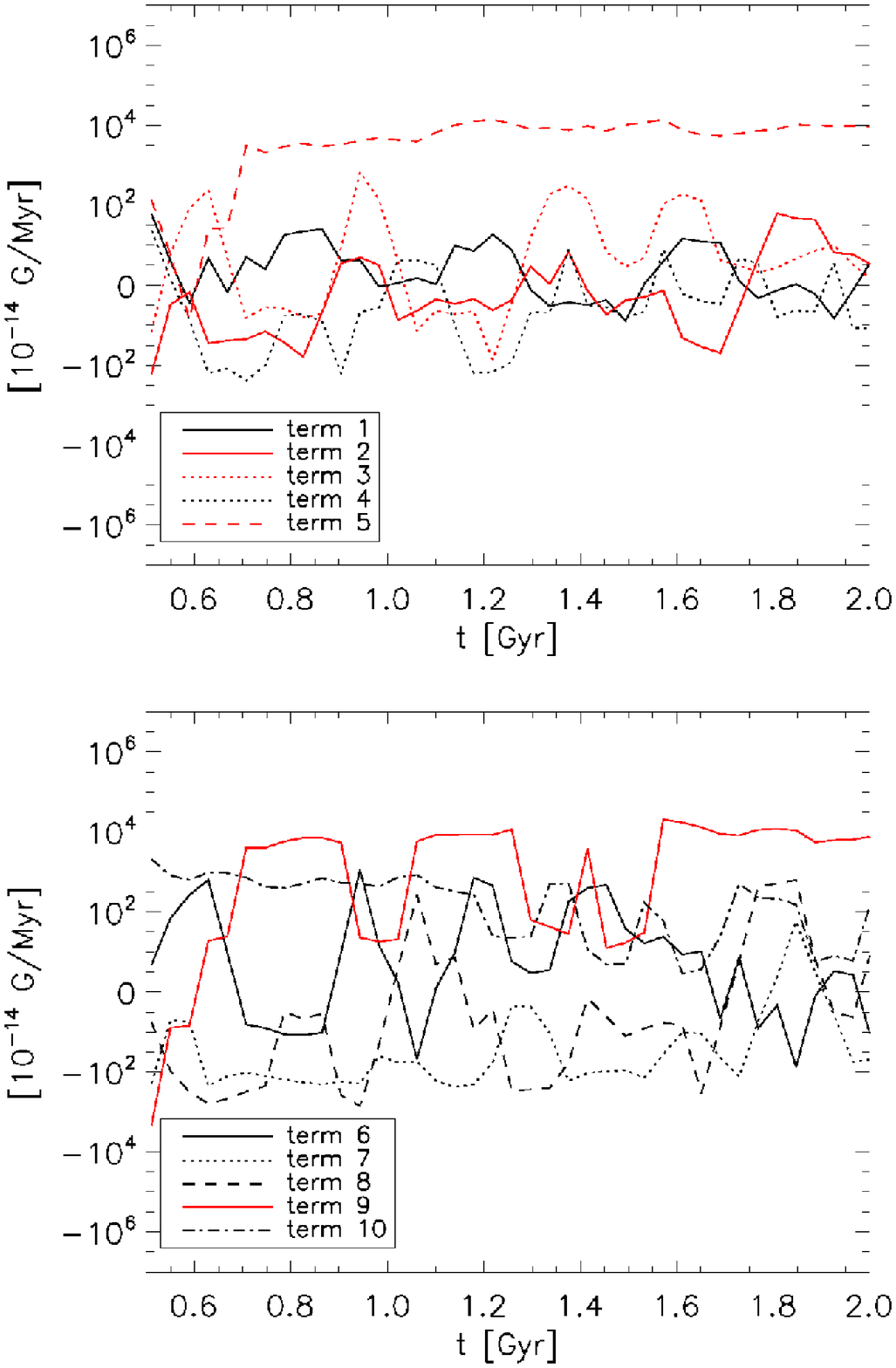, width=0.47\textwidth}
  \caption{Values of the different terms of the induction equation with time  for the normal resolution $\textsc{Gadget}$ simulation using Euler potentials. Upper plot: temporal evolution of the terms responsible for amplification/attenuation of the radial magnetic field (terms 1 to 5). Lower plot: Evolution of terms 6-10, responsible for the evolution of the toroidal magnetic field. Positive values imply amplification, and negative attenuation of the corresponding $\textbf{B}$-component. The non-axisymmetric terms are shown in red, the axisymmetric terms are shown in black.
  }
  {\label{IndEqsAna}}
\end{center}
\end{figure}

In order to quantify the influence of the different terms 1-10 during the simulation we calculated their values in cylindrical bins within the disc (5 to 15 kpc) and their mean value at different times. We have taken the negative values of each term in case of negative magnetic field to distinguish between amplifying and attenuating terms. The result of this calculation is shown in Fig. \ref{IndEqsAna}. The upper plot shows the temporal evolution of the terms responsible for amplification/attenuation of the radial magnetic field (terms 1 to 5) and the lower of the toroidal magnetic field (terms 6 to 10). Positive values imply amplification, and negative attenuation of the corresponding $\textbf{B}$-component. The non-axisymmetric terms are shown in red.

Looking at Fig. \ref{IndEqsAna}, the most important term for the evolution of the radial magnetic field is term 5, i.e. $-\frac{v_\varphi}{r}\frac{\partial B_r}{\partial \varphi}$. Since the toroidal velocity dominates the velocity field, this term is most important although $\frac{\partial B_r}{\partial \varphi}$ is comparatively small. This can be seen following the evolution of the circular velocity and the radial magnetic field more closely: The radial magnetic field is strongest where the circular velocity has its highest value, with a delay of roughly $40$ Myr.
All other terms lie in the same range and therefore compete with each other. Since their values are positive as well as negative, one should not expect a significant amplification on their account. This analysis shows, that even small deviations from axial symmetry are very important for the evolution of the magnetic field in spiral galaxies.

One reaches the same conclusion looking at the terms of the evolution equation for $B_\varphi$. Except for the beginning of the simulation, the leading term here is clearly term 9, $-\frac{v_\varphi}{r}\frac{\partial B_\varphi}{\partial\varphi}$, i.e. the only non-axisymmetric term in this equation. Term 10, which was our candidate for the most important term for axial symmetry, is only the second most important. Both terms depend on the toroidal velocity component, thus demonstrating the importance of the differential rotation for the evolution of the toroidal component of the magnetic field.

Neglecting all non-axisymmetric terms (plotted in red) one finds term 1 ($-B_r\frac{v_r}{r}$) to be largely dominant over term term 4 ($-v_r\frac{\partial B_r}{\partial r}$), in agreement with the theory for the evolution of $B_r$. Also the term responsible for the evolution of $B_\varphi$ is as expected: Term 10 ($-\frac{v_\varphi}{r}B_r$) is the leading term and followed by term 6 ($-B_\varphi\frac{\partial v_r}{\partial r}$). However, term 1 and 6 are both of order $10^{-13}$ G Myr$^{-1}$, thus not being able to account for any significant amplification of our initial magnetic field, and term 10 can only amplify $B_\varphi$ effectively if $B_r$ is amplified.

This behaviour is qualitatively the same also for runs with the direct implementation of the induction equation. We conclude that the non-axisymmetry of the system is the driving force for the observed field amplification in our simulations.

\section{Conclusion and Outlook}
\label{SUMMARY}

We have presented a set of self-consistent simulations of the evolution of magnetic fields in galactic discs performed with the \textsl{N}-body codes $\textsc{Vine}$ and $\textsc{Gadget}$. Hydrodynamics was treated using the SPH method. The evolution of magnetic fields within the framework of ideal MHD was followed by both a direct implementation of the induction equation and a formalism using Euler potentials.

The presented set of simulations shows the importance of a sensible treatment of $\nabla\cdot\textbf{B}$ when simulating magnetic fields in spiral galaxies. Since artificial magnetic monopoles can be responsible for unphysical amplification of the field, more studies of possibilities to avoid or inhibit numerical $\nabla\cdot\textbf{B}$ terms are still needed. Although the description using Euler potentials avoids (physical) magnetic monopoles by construction, the drawback in using them is that they lead to constraints on magnetic helicity. Since helicity fluxes can affect the dynamo process within a mean field theory (\citealp{Brandenburg2005}), Euler potentials would probably not be suitable for simulations including the $\alpha$-effect. Furthermore, Euler potentials do not allow for all initial field configurations, since they are not necessarily single valued and in addition, their derivation can become quite complex. Nevertheless, using topologically simple initial conditions for the magnetic field, the Euler potential formalism seems to be the best tool to follow the ideal evolution of magnetic fields in simulations of spiral galaxies with SPH.

A possible alternative to Euler potentials is the vector potential $\textbf{A}$. The disadvantages are the need for a time integration of $\textbf{A}$ when evolving magnetic fields and the occurrence of second derivatives in the force equation when calculating magnetic forces ($F_\mathrm{mag}\propto\textbf{j}\times\textbf{B}\propto(\nabla\times\textbf{B})\times\textbf{B}
\propto(\nabla\times(\nabla\times\textbf{A}))\times(\nabla\times\textbf{A})$), both leading to lower accuracy in the calculation. On the other hand, the advantages are a somewhat easier derivation of $\textbf{A}$ for a given magnetic field and that there are no constraints on magnetic helicity using a vector potential. It would be definitely interesting to study the differences between simulations utilizing a vector potential and the Euler description, although it could be hard to overcome the problems related to numerical intricacies within a SPH implementation of the vector potential.

The analysis of the different terms of the induction equation applied to our simulations clearly show that the non-axisymmetry of the velocity and magnetic field cannot be ignored in any consideration of the kinematic dynamo. There are two main processes leading to angular momentum transport and hence non-axisymmetry in spiral galaxies: Internal driving due to spiral structure and bar formation (the former considered in the presented paper) and external driving due to interaction with other galaxies. Simulations of interacting systems would therefore enrich our understandings even further on how large scale magnetic fields evolve due to large scale velocity fields.

Our simulations show only a weak amplification of the initial magnetic field. Observations of spirals galaxies at high redshifts suggest that their magnetic field strengths were at least as strong as the magnetic fields at the current epoch within few Gyrs of the Big Bang (\citealp{Kronberg2007}). Assuming initial strengths of order $B_\mathrm{IGM}=10^{-9}$ G an amplifying process should therefore account for four orders of magnitude of increase within few Gyrs in order to reach the observed values of $\approx 10 \mu$G. Since our simulations of a purely kinematic dynamo account at best for one order of magnitude, there is still need for a more complete scenario with additional subgrid physics. Such subgrid physics should comprise the $\alpha$-effect due to turbulent gas motions below the resolution limit, estimated from local high-resolution MHD simulations and observations of turbulent motions in nearby galaxies. Hereby, potentially the most promising ansatz is the cosmic ray driven dynamo (\citealp{Lesch&Hanasz2003}, \citealp{Hanasz&Lesch2005}). Given the fact, that the presented simulations reveal the complete three dimensional velocity field to fully account for the large-scale structure of the magnetic field, we believe that \textsl{N}-body SPH together with sensible subgrid physics will be apt to test our understanding of the evolution of magnetic fields in spiral galaxies numerically.

\section*{Acknowledgments}
We thank the referee Daniel Price for numerous very helpful comments and suggestions which improved the paper a lot.
HK is grateful
for interesting discussions with Michal Hanasz, Daniel Price and Axel Brandenburg.
The authors also greatly thank Rainer Beck for providing Fig. \ref{M51}.
This research was supported by the DFG cluster of excellence `Origin and
Structure of the Universe' (www.universe-cluster.de).

\bibliographystyle{mn2e}
\bibliography{MagneticFieldStructure}

\label{lastpage}
\end{document}